%% file: kimianet.tex
\documentclass[journal]{IEEEtran}

\usepackage{natbib}
\usepackage{graphicx}
\usepackage{algorithm}
\usepackage[noend]{algpseudocode}
\usepackage[table,xcdraw]{xcolor}
\usepackage{booktabs}
\usepackage{fancyhdr}
\usepackage{authblk}

\begin{document}
\title{Fine-Tuning and Training of DenseNet for Histopathology Image Representation\\ Using TCGA Diagnostic Slides}

\input{Sections/authors}

\IEEEtitleabstractindextext{%
\begin{abstract}
\input{Sections/abstract}
\end{abstract}

\begin{IEEEkeywords}
Histopathology, Deep Learning, Transfer Learning, Image Search, Image Classification, Deep Features, Image Representation, TCGA
\end{IEEEkeywords}}

\maketitle

\IEEEdisplaynontitleabstractindextext
\IEEEpeerreviewmaketitle

\input{Sections/Introduction}

\input{Sections/Literature_Review}

\input{Sections/KimiaNet-Data_and_Training.tex}

\input{Sections/Experiments}

\input{Sections/Summary_and_Conclusions}

\ifCLASSOPTIONcaptionsoff
  \newpage
\fi

\bibliographystyle{bibstyle}  
\bibliography{references}

\section{Appendix}

  \input{Tables/tab-diagnosis-abbrv}

\end{document}

%% file: Sections/authors.tex
\author[1]{Abtin Riasatian}
\author[1]{Morteza Babaie}
\author[1]{Danial Maleki}
\author[1]{Shivam Kalra}
\author[2]{Mojtaba Valipour}
\author[1]{Sobhan Hemati}
\author[1]{Manit Zaveri}
\author[1]{Amir Safarpoor}
\author[1]{Sobhan Shafiei}
\author[1]{Mehdi Afshari}
\author[1]{Maral Rasoolijaberi}
\author[1]{Milad Sikaroudi}
\author[1]{Mohd Adnan}
\author[3]{Sultaan Shah}
\author[3]{Charles Choi}
\author[3]{Savvas Damaskinos}
\author[4]{Clinton JV Campbell}
\author[5]{Phedias Diamandis}
\author[6]{Liron Pantanowitz}
\author[1]{Hany Kashani}
\author[2, 7]{Ali Ghodsi}
\author[1, 7]{H.R. Tizhoosh \thanks{Manuscript submitted for publication on December 31, 2019. Corresponding authors: Morteza Babaie (email: mbabaie@uwaterloo.ca), H.R. Tizhoosh (email: tizhoosh@uwaterloo.ca)}}

\affil[1]{Kimia Lab, University of Waterloo, 200 University Ave. W., Waterloo, ON, Canada}
\affil[2]{School of Computer Science, University of Waterloo, 200 University Ave. W., Waterloo, ON, Canada}
\affil[3]{Huron Digital Pathology, 1620 King Street North, St. Jacobs, ON, Canada}
\affil[4]{Department of Pathology and Molecular Medicine, McMaster University, Hamilton, Canada}
\affil[5]{Laboratory Medicine and Pathobiology, University of Toronto, ON, Canada}
\affil[6]{Department of Pathology, University of Pittsburgh Medical Center, PA, USA}
\affil[7]{Vector Institute, 661 University Ave Suite 710, Toronto, ON, Canada}

%% file: Sections/abstract.tex
Feature vectors provided by pre-trained deep artificial neural networks have become a dominant source for image representation in recent literature. Their contribution to the performance of image analysis can be improved through fine-tuning. As an ultimate solution, one might even train a deep network from scratch with the domain-relevant images, a highly desirable option which is generally impeded in pathology by lack of labeled images and the computational expense. In this study, we propose a new network, namely \emph{KimiaNet}, that employs the topology of the DenseNet with four dense blocks, fine-tuned and trained with histopathology images in different configurations. We used more than 240,000 image patches with $1000\times 1000$ pixels acquired at 20$\times$ magnification through our proposed ``\emph{high-cellularity mosaic}'' approach to enable the usage of weak labels of 7,126 whole slide images of formalin-fixed paraffin-embedded human pathology samples publicly available through the The Cancer Genome Atlas (TCGA) repository. We tested KimiaNet using three public datasets, namely TCGA, endometrial cancer images, and colorectal cancer images by evaluating the performance of search and classification when corresponding features of different networks are used for image representation. As well, we designed and trained multiple convolutional batch-normalized ReLU (CBR) networks. The results show that KimiaNet provides superior results compared to the original DenseNet and smaller CBR networks when used as feature extractor to represent histopathology images.  

%% file: Sections/Introduction.tex
\section{Introduction}
Conventional light microscopy is a well-established technology with centuries of history. The adoption of digital pathology that replaces the microscope with a digital scanner and computer monitor has gained momentum in recent years. The process of digitization of whole slide images (WSIs) offers many advantages such as more efficient workflows, easier collaboration and telepathology, and new biological insights into histopathology data through usage of image processing and computer vision algorithms to detect relevant clinicopathologic patterns \citep{gurcan2009histopathological,tizhoosh2018artificial}. 

The recent progress in machine learning, particularly deep learning, provides a major argument for proponents of modern pathology to justify the benefits of going digital. Pre-trained, fine-tuned and de novo trained deep networks are being proposed for diverse classification, prediction and retrieval tasks. However, processing WSIs, with or without machine learning, has its own challenges \citep{madabhushi2009digital}. WSIs of histopathology are large files containing complex histologic patterns. Hence, compact and expressive image representation, which is fundamental to image analysis in pathology presents many challenges. Handcrafted features, i.e., image descriptors that have been manually designed based on general image processing knowledge, have been in use as a solution for several decades \citep{jegou2011aggregating}. Many studies, however, have shown that deep features, i.e., high-level embeddings in a properly trained deep network, can outperform handcrafted features in most applications \citep{kumar2017comparative}. As a result, many different convolutional  architectures have been trained and introduced to provide features either directly or through transfer learning \citep{shin2016deep,mormont2018comparison}. This has created new questions in computer vision research, and consequently in medical image analysis as to which network topology is most suitable for a given task. Is transfer learning sufficient to solve specific problems? What challenges and benefits does training an entire network from scratch entail? \citep{niazi2019digital,yu2016discriminative,burt2018deep}.    

In this work, we focus on one of the most successful convolutional topolgies in the literature, namely the DenseNet \citep{huang2017densely,zhang2019full,zhang2019multiple}. We fine-tuned and trained the DenseNet architecture with a large number of histopathology patches at 20 times magnification (20$\times$) and compare its search and classification performance, as a feature extractor, against the original DeneseNet trained with 1.2 million natural images from ImageNet \citep{deng2009imagenet}. While it is generally expected that fine-tuning and training should deliver more accurate results than transfer learning, the data and experimental challenges inherent in this may easily prevent the expected effects \citep{glorot2010understanding,najafabadi2015deep,srivastava2015training}.

In the following sections, we briefly review the relevant literature. We describe how we fine-tuned and re-trained a network, using a large public dataset, which we have named KimiaNet. We report the results of applying KimiaNet on three public datasets for search and classification. We provide details of all experiments to demonstrate the superiority of the KimiaNet features over the original DenseNet features. As well, we report the details of four smaller networks that we designed and trained for bench-marking against KimiaNet.  

%% file: Sections/Literature_Review.tex
\section{Literature Review}
The literature on the applications of deep learning in digital pathology is diverse and contains a multitude of approaches \citep{janowczyk2016deep,niazi2019digital,campanella2019clinical}. As we are focused on image representation, in this section we only review recent works that have used, fine-tuned or trained deep networks for different purposes in digital pathology \citep{janowczyk2016deep}. 

\textbf{Pre-Trained Networks --}
Two early examples of off-the-shelf feature extractors are  Overfeat \citep{sharif2014cnn} and DeCaf \citep{donahue2014decaf} used successfully in breast cancer classification by capturing and combining different fully connected (FC) layers \citep{8122889}. Combining Inception (V3) features (extracted from multi-magnification pathology images) with a fully connected layer has been used for binary classification in breast cancer metastasis analysis \citep{liu2017detecting}.

 Feature selection from pre-trained deep networks has also been performed in many pathology tasks like HEp-2 protein classification \citep{phan2016transfer}. Using features from pre-trained networks on pathology domains is another way to classify the pathology problems \citep{mormont2018comparison}. For instance, features from DeepLoc \citep{almagro2017deeploc} have been used to classify  protein subcellular localization \citep{kraus2017automated}. As well, deep features from pre-trained networks have been successfully used for image search \citep{kalra2019yottixel,kalra2019pan}.

\textbf{Fine-Tuned Networks --}
It is generally expected that fine-tuning should increase the accuracy of a pre-trained network or the expressiveness of its features in pathology image classification \citep{kieffer2017convolutional}. \cite{faust2018visualizing} fine-tuned the last 2 convolutional blocks of a VGG-19 model initialized with ImageNet weights. More specifically, they utilized annotated patches of size $1024$ by $1024$ at $20 \times$ magnification from a neuropathology dataset with 13 distinct classes for transfer learning. Eventually, they reported a $95\%$ accuracy for the classification task. \cite{faust2019intelligent} also fine-tuned a VGG-19 model utilizing an extended version of the latter dataset with 74 distinct classes annotated employing the World Health Organization (WHO) tomour classification scheme. Ultimately, they reported the training validation accuracy of $66\%$ over the images spanning the 74 trained classes.

\textbf{Trained Networks --}  \cite{coudray2018classification} trained an Inception V3 model on $1,176$ lung WSIs from the TCGA dataset. They extracted patches of size $512\!\times\!512$ pixels from Lung Adenocarcinoma (LUAD), Lung Squamous Cell Carcinoma (LUSC), and  healthy lung tissue slides. Subsequently, they predicted diagnosis associated with any given WSI by assessing the majority vote among the patch level predictions of all tiles of the same slide. Finally, they reported around $97\%$ accuracy for both $20 \times$ and $5 \times$ magnifications. 

\cite{fu2019pan} employed transfer learning on $17,396$ fresh-frozen tissue images from the TCGA dataset. Their dataset contained $42$ classes, including $28$ tomour types and $14$ healthy tissues. As they removed the non-informative tiles with minimal gradient magnitude, they fine-tuned an Inception V4 utilizing $6.5$ million patches of size $512 \times 512$ pixels at $20 \times$ magnification extracted from the training set WSIs. In the end, they evaluated the power of the learned visual representation by investigating the connection between the WSIs and genome information. 

\cite{wei2019pathologist} trained a ResNet architecture using five pathological and benign patterns of lung parenchyma, annotated by three pathologists. They also reported a high level of agreement between their model outcome and the pathologists final diagnosis.

\cite{campanella2019clinical} suggested a framework based on multiple instance learning and deep neural networks to evaluate WSI-level diagnosis to avoid time-consuming and expensive annotations. First, they trained several ResNet models on different subsets of a dataset made of $44,732$ private slides to learn visual representation at both $20 \times$ and $5 \times$ magnifications. The dataset included prostate cancer, basal cell carcinoma, and metastatic breast cancer involving the lymph nodes. Next, they chose a subset of most the suspicious tiles from each WSI to pass it through a Recurrent Neural Network (RNN) to predict the WSI-level prediction. \cite{bilaloglu2019efficient} trained a deep convolutional neural network called PathCNN on a dataset comprised of lung cancer, kidney cancer, breast cancer, and non-neoplastic tissues from the TCGA repository. In particular, they extracted regions of size $512 \times 512$ pixels at $20 \times$ magnification and then downsampled them to $299
\times 299$ pixel patches. Similar to previous studies, they predicted WSI-level diagnosis based on the aggregation of the patch-level prediction.

There are a multitude of other possible approaches to customizing deep solution for histopathology images. Multi-instance learning \citep{campanella2019clinical}, teacher-student learning \citep{watanabe2017student}, and visual dictionaries \citep{zhu2018multiple} are among most investigated, to mention a few. Here, we are mainly focused on training and fine-tuning of most commonly used deep topologies. 

\vspace{0.1in}
As the literature shows, we still need to investigate the effect of fine-tuning and training from scratch on expected performance improvement specially using deep embeddings as image features for various tasks. This should ideally be performed using a large public dataset with raw heterogeneous cases not particularly curated for training, hence easily repeatable on many other repositories. As well, images should be processed at high magnification (e.g., $20 \times$ or higher) and large enough patches (e.g., $500 \times 500 \mu m^2 \sim 1000 \times 1000$ pixels) to model and cover the workflow for most diagnostic cases. We will investigate the fine-tuning and training of such a network and test its features for search and classification using three public datasets.

Although proposing new architectures and learning algorithms have been the main vehicle of progress within the AI community in recent years, customizing existing topologies is absolutely necessary and may still not be easily possible due to both data and computational challenges. The TCGA repository, for instance, is large and publicly available. However, due to the absence of pixel-level and regional labels of gigapixel files, it is not readily available for training deep networks. Solving practical challenges like this to exploit the discriminative power of deep networks appear to also  be a valuable way of knowledge creation. The contribution of the proposed high-cellularity mosaic as whole-slide image representation is enabling the usage of the unlabelled image data by providing an implicit regional annotation, i.e., the mosaic patches of high-cellularity.

Our main contribution is to propose a specialized patch selection method to create a high-cellularity collection -- called \emph{cellMosaic} -- to enable the usage of weak WSI-level labels of a public multi-organ cancer image archive at high magnification with high-resolution  patches with no downsampling for training of a densely connected network. The proposed KimiaNet can then be employed for feature extraction in histopathology.

%% file: Sections/KimiaNet-Data_and_Training.tex
\section{KimiaNet - Data and Training}
We name our network ``\emph{Kimia}Net''\footnote{The Persian/Arabic word \emph{kīmiyā}  and its Greek version \emph{khēmeía} appear to originate from the Coptic word \emph{kēme} (meaning \emph{Egypt}) and is believed to be the root word for ``chemistry'' and associated with the  \emph{alchemy} that tried to purify metals and convert them to gold.} as we are convinced that the actual information hidden in big image data can only be extracted through extensive fine-tuning, or better, training from scratch. As there have been extensive research on network topologies, we have chosen a dependable network, called \emph{DenseNet}, as the basis for our investigations. 

Customizing well-established architectures for specific and sensitive tasks not only appear to be justified but also seem to be necessary for the sake of application-oriented categorization and end-user awareness. One  example is certainly the ``CheXNet'' that is a fine-tuned DenseNet using x-ray images \citep{rajpurkar2017chexnet}.

\begin{figure*}[htb]
\centering
\includegraphics[width=0.85\textwidth]{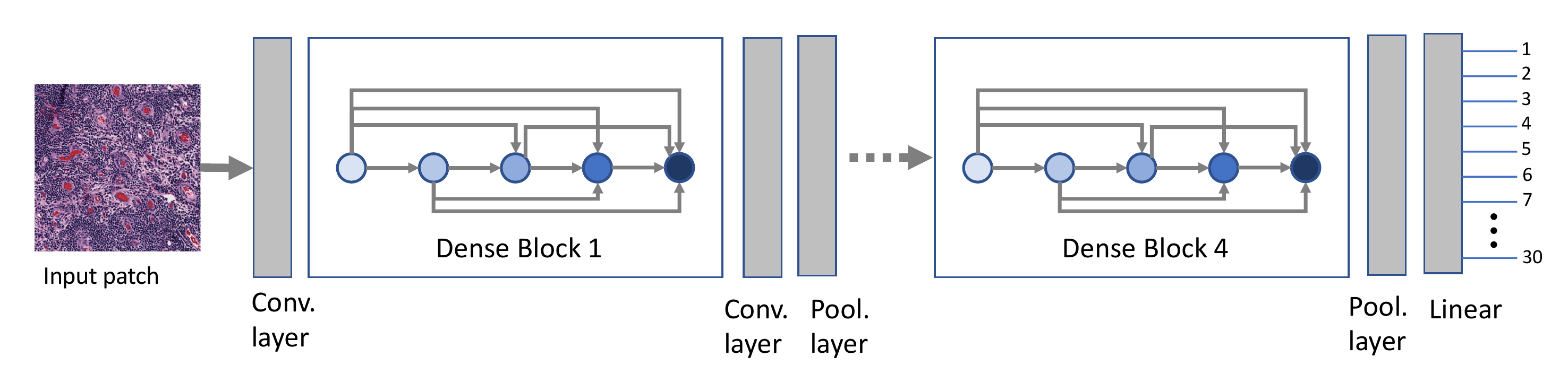}
\caption{DenseNet architecture of KimiaNet.}
\label{fig:densenet}
\end{figure*}

Based on the observation that convolutional networks can
be very deep, yet more accurate, and still efficient to train ``if they contain shorter connections between layers close to the input and those close to the output'', Gao Huang and his colleagues introduced the densely connected convolutional networks \citep{huang2017densely}. The network consists of several dense blocks with preceding convolutional and pooling layer (Figure \ref{fig:densenet}). Most recent works refer to DenseNet topology as a reliable candidate solution for image representation in histopathology \citep{campanella2019clinical}. Beside the popularity of DenseNet \citep{lee2017can, liu2018deep}, we have already experimented with its features in our previous works \citep{babaie2019deep, kalra2019yottixel,kalra2019pan}. In addition, compared to the top-10 matching evaluation of networks for bench-marking through ImageNet, DenseNet is certainly a compact architecture with a smaller footprint; the size of DenseNet-121 topology with almost 7M parameters amounts to almost 10\% of EfficientNet-B7 (66M parameters) and 0.8\% of FixResNeXt-101 $32\times$48d (829M parameters)\footnote{https://sotabench.com/benchmarks/image-classification-on-imagenet}.

\subsection{Public Image Datasets}
It is paramount to use public data such that results are reproducible by other researchers. We downloaded and used three public image datasets: pan-cancer images from The Cancer Genome Atlas (TCGA) repository ($\approx$ 33,000 WSIs for 32 primary diagnosis), endometrial cancer images ($\approx$ 3,300 patches from 4 classes), and colorectal cancer images (5,000 patches from 8 classes). Whereas TCGA provides WSIs with primary diagnosis for the entire image, both colorectal and endometrial datasets contain labelled patches. We used most TCGA images for training and some for testing; both colorectal and endometrial datasets were exclusively used for testing. 

\subsubsection{TCGA Images}
The TCGA repository (i.e., Genomic Data Commons, GDC\footnote{https://portal.gdc.cancer.gov/}) with 30,072 WSIs is a publicly available repository \citep{gutman2013cancer,tomczak2015cancer,cooper2018pancancer}. 
We recorded 29,120 WSI fully readable files at 20$\times$ magnification (approximately 6 terabytes in compressed form) to prepare the dataset for training. Although 40$\times$ magnification images were also available in many cases, we did use 20$\times$ magnification to maximize the size of the dataset. The dataset contains 25 anatomic sites with 32 cancer subtypes. Brain, endocrine, gastrointestinal tract, gynecological, hematopoietic, liver/pancreaticobiliary, melanocytic, prostate/testis, pulmonary, and urinary tract had more than one primary diagnoses such that they could be used for subtype classification. From the 29,120 WSIs, 26,564 specimens were neoplasms, and 2,556 were non-neoplastic. A total of 17,425 files comprised of frozen section digital slides were removed from the dataset due to their lower quality. We did not use frozen sections because the freezing artefacts in these images can ``confound routine pathological examination or image analysis algorithms'' \citep{cooper2018pancancer}. We kept 11,579 of permanent hematoxylin and eosin (H\&E) sections for training and testing. 
We did not remove manual pen markings from the slides when present. 
This pre-selection of TCGA images will be further refined to assemble the training, validation and testing datasets (see Section \ref{sec:traindata}).

\subsubsection{Endometrium dataset}
Recently the endometrium dataset was introduced to compare a deep learning method (HIENet) classification ability versus four experienced pathologists \citep{8854180}. In this dataset, there are four classes of endometrial tissue, namely normal, endometrial polyp, endometrial hyperplasia, and endometrial adenocarcinoma. Table \ref{Endometrial-dataset} shows the class distribution of all $3,302$ images in the endometrium dataset. Patches of size $640\!\times\!480$ pixels are extracted from $20 \times$ or $10 \times$ magnification WSIs and saved as JPEG files\footnote{Download: https://doi.org/10.6084/m9.figshare.7306361.v2}. Although all slides have been prepared and scanned at the same hospital, considerable stain variation can be observed across endometrium patches.

\begin{table}[t]
\caption{Endometrium dataset \citep{8854180}}
\label{Endometrial-dataset}
\centering
\resizebox{0.4\textwidth}{!}{%
\begin{tabular}{l|c}
Class & Number of patches \\ \hline\hline
Normal Endometrium (NE) & 1,333 \\ \hline
Endometrial Polyp (EP) & 636 \\\hline
Endometrial Hyperplasia (EH) & 798 \\ \hline
Endometrial Adenocarcinoma (EA) & 535 \\ \hline
\end{tabular}%
}
\end{table}

\subsubsection{Colorectal Cancer Dataset}
One of the first digital pathology classification datasets, the colorectal cancer images \citep{kather2016multi}, consists of 5,000 samples in 8 classes and 625 small patches ($150\!\times\!150$ pixels) in each class. Labels in this dataset are tumour epithelium, simple stroma, complex stroma, immune cells, debris, normal mucosal glands, adipose tissue and background patches. 

\begin{figure}[t]
\centerline{\includegraphics[width=0.5\textwidth]{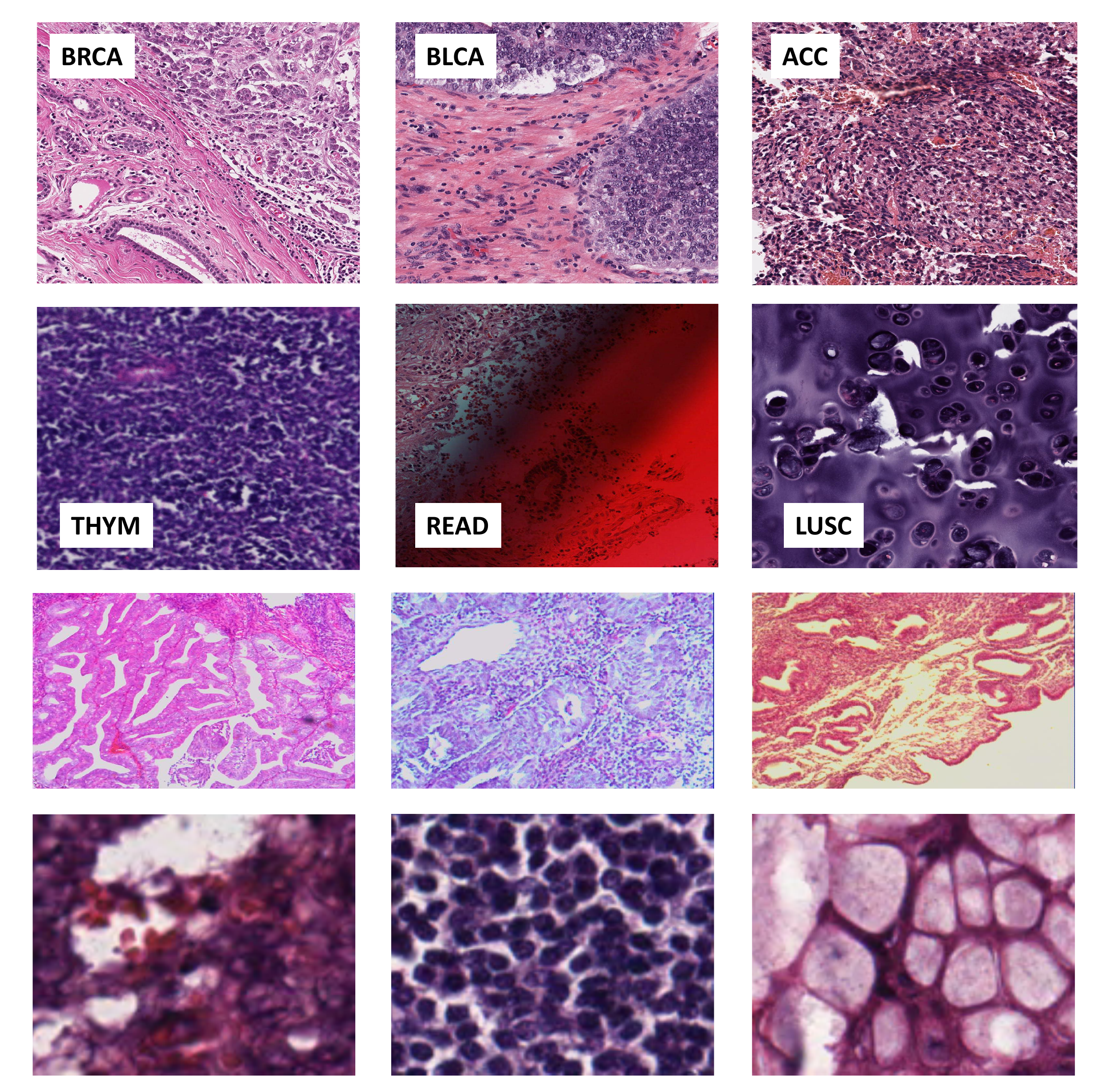}}
\caption{Sample patches from the three datasets: TCGA repository (top two rows with the 1st row for good quality and 2nd row for low quality samples), endometrial cancer (3rd row), and colorectal cancer (bottom row).}
\label{fig:endometrium}
\end{figure}

\vspace{0.1in}
Figure \ref{fig:endometrium} shows sample images for each of the three datasets. 

\subsection{Processing Unlabelled Big WSI Data}
Due to the large size of digital pathology images, representing WSI files is still an obstacle for many tasks in computational pathology \citep{tizhoosh2018artificial}. Most approaches therefore focus on patch processing. The TCGA data only provides WSIs such that patches have to be extracted. As well, images are not labelled in the common sense. That means there is no manual delineation of regions of interest (i.e., malignant pixels); TCGA WSIs are associated with a primary diagnosis for the entire image which may also contain healthy tissue. This makes the creation of a dataset of patches somewhat difficult as we need to feed labelled patches (i.e., small sub-images) into a deep network.

Yottixel is a recently proposed image search engine for histopathology \citep{kalra2019yottixel,kalra2019pan}. We used a modified Yottixel indexing followed by post-processing to extract labelled patches from TCGA dataset (see Algorithm \ref{alg:yottixel}). Yottixel assembles a ``mosaic'' of each WSI at magnification $m_I$ through patching and clustering at magnification $m_C$ with patches at size $l\times l$ grouped in $n_C$ clusters. The mosaic is a rather small collection of patches, i.e., $p$ percentage of the image area, at magnification $m_I$ to represent the entire WSI. Here, the main difference with the original Yottixel mosaic is the function \emph{cellMosaic} (line 13, Algorithm \ref{alg:yottixel}) that modifies the mosaic $M$ into a new mosaic $M'$ by removing all patches with low cellularity by taking the top $T_\textrm{Cell}$ percent of cellularity-sorted patches. Based on the assumption that many high-grade carcinomas may have higher cellularity levels compared with healthy tissue, $M'$ enables us to use the WSI label (i.e., the primary diagnosis) for all remaining patches of each WSI, hence making the TCGA data usable for training a network.

\begin{figure*}[htb]
\centerline{\includegraphics[width=\textwidth]{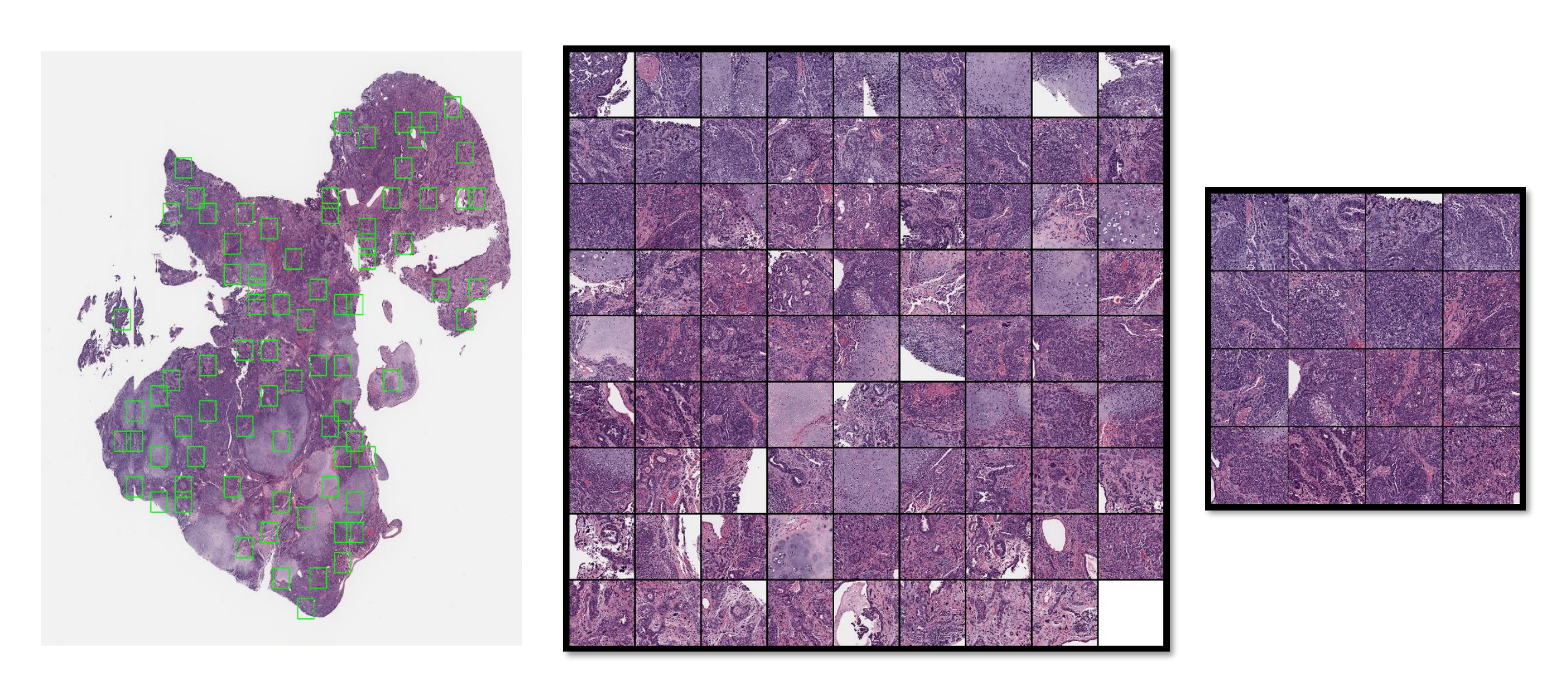}}
\caption{A WSI and its selected mosaic patches (left), Yottixel mosaic with 80 patches (middle), modified \emph{cellMosaic} with 16 patches (right).}
\label{mosaic_figure}
\end{figure*}

\begin{algorithm}[t]
\caption{Modified Yottixel Algorithm} \label{alg:yottixel}
\begin{algorithmic}[1]
\State $m_I \gets 20x$ \Comment{Magnification for indexing}
\State $m_C \gets 5x$ \Comment{Magnification for clustering}
\State $l \gets 1000$ \Comment{Patch size $l\times l$ at $m_I$}
\State $n_C \gets 9$ \Comment{Number of clusters at $m_C$}
\State $p \gets 15\%$ \Comment{Mosaic percentage}
\State $T_\textrm{Cell} \gets 20\%$ \Comment{Top cases among sorted cellularity}
\State $\mathbf{A} \gets \textrm{readWSI}(fileName)$ \Comment{Read an image}
\Procedure{YottixelIndex}{$\mathbf{A},m_I,m_C,l,n_C,p,T_\textrm{Cell}$}
\State $\mathbf{S}\gets \textrm{Segment}(\mathbf{A},m_C)$ \Comment{Separate  tissue/background}
\State $ P\gets \textrm{Patching}(\mathbf{A},\mathbf{S},m_I,l)$ 
\Comment{Get all patches}
\State $C \gets \textrm{KMeansCluster}(P)$ \Comment{Cluster patches}
\State $M \gets \textrm{getMosaic}(C,p,\mathbf{A})$ \Comment{Select a mosaic}
\State $M' \gets \textrm{cellMosaic}(M, T_\textrm{Cell})$ \Comment{Keep cell patches}
\State $F \gets \textrm{Network}(M')$ \Comment{Get features}
\State \textbf{return} $F$\Comment{Set of features for $\mathbf{A}$}
\EndProcedure
\end{algorithmic}
\end{algorithm}

\subsection{Training Data}
\label{sec:traindata}
To create the dataset, we first eliminated frozen sections so that only permanent section diagnostic slides were left. The low quality of frozen section images might negatively affect training. In order to create a versatile dataset, we divided the data into groups with most detailed labels, each unique group being specified by combination of ‘morphology’, ‘primary diagnosis’ and  ‘tissue or organ of origin’ label. Then, we removed the groups with fewer than 20 cases so the dataset can be used at the most detailed level by specifying the label of each class with the mentioned combination; hence, each class has at least 2 cases (ten percent of 20) test samples. For example, one of the deleted groups was ['8020/3', 'Carcinoma, undifferentiated, NOS', 'Tail of pancreas'] which had only one case. As a result of this process, 2 of the 32 classes of primary diagnoses were removed, the UCEC (Uterine   Corpus   Endometrial Carcinoma) class due to the morphology information not being reported at the time of creating the dataset and the DLBC (Lymphoid  Neoplasm  Diffuse Large B-cell Lymphoma) class for not having any detailed groups with at least 20 cases. We used tomour type categorization based on established literature \citep{cooper2018pancancer}. For instance, LUAD (lung adenocarcinoma), LUSC (lung squamous cell carcinoma) and MESO (mesothelioma) would all fall under ``Pulmonary Tumours''. This is particularly useful for evaluation of horizontal search.
The test and validation datasets were chosen from the single cases randomly selecting from 10\% of WSIs within each class. The rest of the slides were assigned to the training dataset. All cases that did not contain diagnostic or morphological information were removed from all datasets. This resulted into a test dataset of 777 slides, a validation dataset of 776 slides and training dataset of 7,375 diagnostic slides. All three sets are disjoint. In Particular, we selected both test and validation sets to only consist of those patients with only one diagnostic slide. As well,  WSIs with no magnification information or with magnification lower than 20$\times$ were removed. This led to creation of a test dataset of 744 slides, a validation dataset of 741 slides and a training dataset of 7,126 slides (\textbf{a total of 8,611 WSIs}). Extracting $500\mu\times 500\mu$ at 20$\times$ finally resulted in 1,198,118 patches for training, 121,801 patches for validation, and 116,088 patches for testing.

Pathologists  generally use different patch sizes and samples to find different types of information. For instance, 10$\times$ is used for gross features and infiltrates,  20$\times$ for more detailed histology patterns, and 40$\times$ for fine nuclear and cellular details. Both magnification level and patch size are set based on empirical evidence and computational convenience (some works have used similar settings as in this work, e.g., see \cite{faust2018visualizing}). The algorithms proposed in this work can be run for any magnification and any patch size if desired histologic features are apparent and/or computational resources available.

\begin{figure}[t]
\includegraphics[width=0.99\columnwidth]{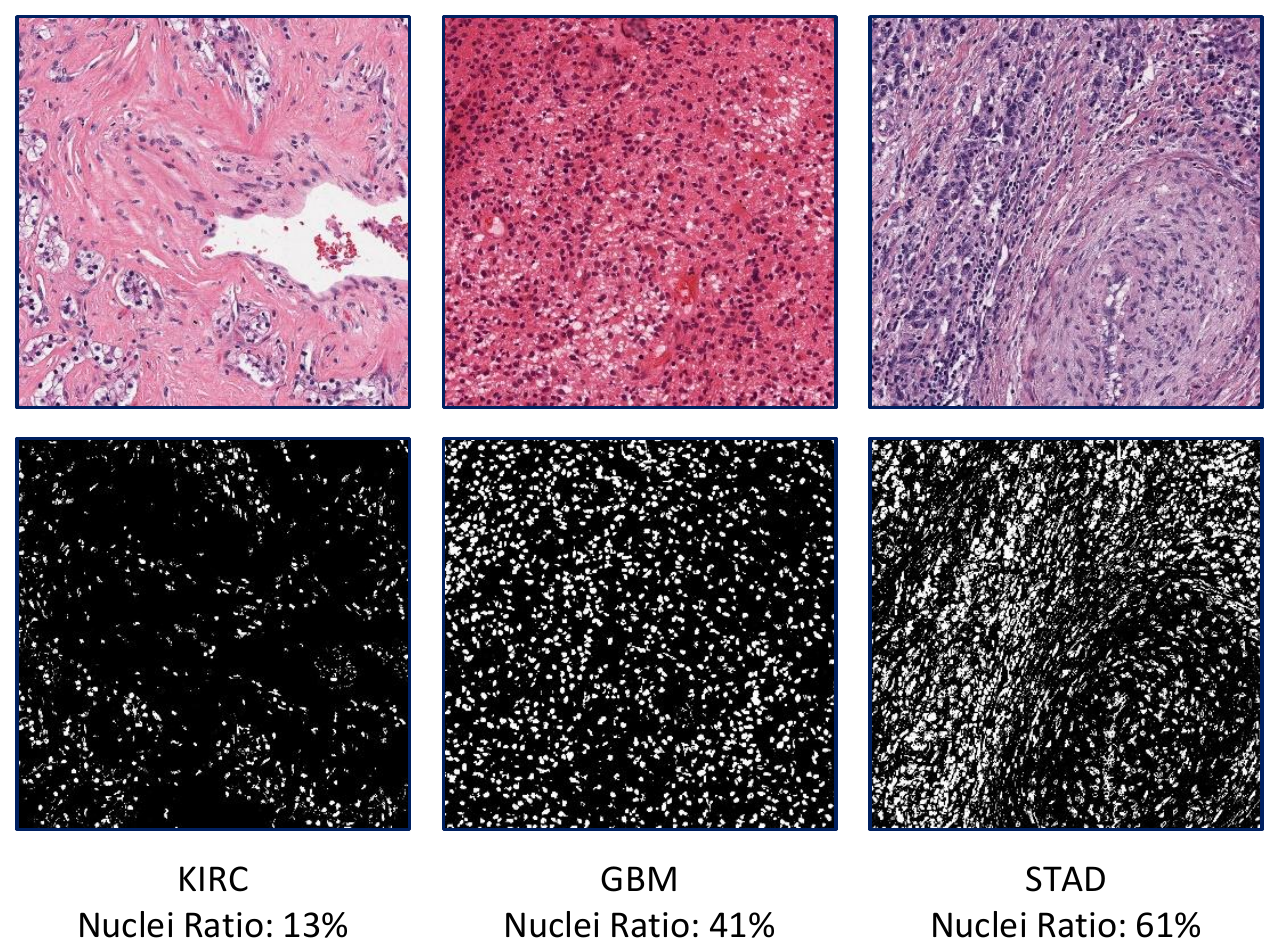}
\caption{Samples for cell segmentation. }
\label{fig:sampleCellSegments}
\end{figure}

 The dataset was still not suitable for training; as the WSIs were labeled with diagnosis, and not the patches. However, we intended to train the network at the patch level. The patch dataset included multiple healthy/benign patches associated with a diseased WSI. This would perhaps confuse any deep network during training. As carcinomas are generally associated with uncontrolled cell growth, mainly embodied in areas with unusually high presence of cell nuclei (e.g., small cell carcinoma is extremely hypercellular), cellularity of patches can be used to eliminate most benign/healthy patches \citep{travis2014pathology}\footnote{Although abnormal and disrupted tissue architecture is another major criterion to select candidate patches, this may be more challenging to accomplish compared to cellularity measurements.}. We chose hypercellularity to automate patch selection as this is one of the principal hallmark features of cancer that spans most neoplasms.  Cellularity can be used as an initial filter to select patches with a higher probability of malignancy. However, we realize that reactive or inflammatory tissue features may in some cases also be included in this set. While we do recognize that non-neoplastic cell types may show hypercellular regions, we feel these caveats are outweighed by the superior automation provided by this approach. Similar approaches have been used recently when patches with minimal gradient magnitudes were eliminated \citep{fu2019pan}. Hence, we measured the \emph{cellularity} of each patch. This was done by first deconvolving the patch color from RGB to hematoxylin and eosin channels using color deconvolution \citep{onder2014review}\footnote{We used the function available in HistomicsTK library on GitHub: The RGB image is  transformed into optical density space, and then projected onto the stain vectors in the columns of the stain matrix $W$, a 3$\times$3 matrix containing the color vectors in columns. For two stain images the third column is zero and will be complemented using cross-product. For deconvolving H\&E stained image, $W$ is set to [[0.650, 0.072, 0], [0.704, 0.990, 0], [0.286, 0.105, 0]].}. Then a binary mask was created from the hematoxylin channel using a constant threshold, set empirically, to get the cellularity ratio for each patch (Figure \ref{fig:sampleCellSegments}). Finally, training and validation dataset patches were each sorted with respect to their ratio of cellularity (number of pixels with value 1 in the created mask over the number of all patch pixels). The top $T_\textrm{Cell}=20\%$ patches (lines 6 and 13, Algorithm\ref{alg:yottixel}) with regard to their sorted ratio were selected with the additional constraint of having a file size larger than a specific threshold (e.g., $>$100 KB). This resulted into final training and validation datasets containing \textbf{242,202}  and \textbf{24,646} patches, from \textbf{7,126 and 741 WSIs,} respectively. Hence, we are using a 80\%-10\%-10\% split for training/validation/testing to assign as many samples as possible to the training data. Besides, the restrictions on testing data geared the ratio also toward larger this split.

\subsection{Training}
We run the training/fine-tuning in several configurations to generate KimiaNet-I (last DenseNet-121 block trained), KimiaNet-II (last two DenseNet-121 blocks trained), KimiaNet-III (last three  DenseNet-121 blocks trained), and KimiaNet-IV (all DenseNet-121 blocks trained).

We used the \emph{PyTorch} platform to train and test the models described above. We trained each model on 4 Tesla V100 GPUs with 32GB memory per GPU. We set the batch size of 256, 128, 128 and 64 for KimiaNet-I, KimiaNet-II, KimiaNet-III and KimiaNet-IV,  respectively. Each network was trained for roughly 20 epochs. Each epoch took approximately 60, 75, 90 and 110 minutes for KimiaNet configurations I,II,III and IV, respectively. We used 30 classes of primary diagnoses as the KimiaNet output (see Table \ref{tab:diagnosis-abbrv} in Appendix). 
The stopping criterion for training is a decrease in validation accuracy for three consecutive epochs.
We used the Adam optimizer \citep{kingma2014adam} for optimization of our models with an initial learning rate of 0.0001 and scheduler that decreases the learning rate every 5 epochs. We set the  cross-entropy as the loss function for our models to measure the performance of the classification model. For each of the models, we used the ImageNet pre-trained weights for initialization. One has to bear in mind that the DenseNet has been trained with 1.2 million images for 1,000 classes; KimiaNet has been trained with 242,202 images (data ratio $=242,202/1,200,000\approx0.20$ ) for 30 classes (class ratio $= 30/1,000=.03$). 

\begin{figure}[t]
\centerline{\includegraphics[width=0.85\columnwidth]{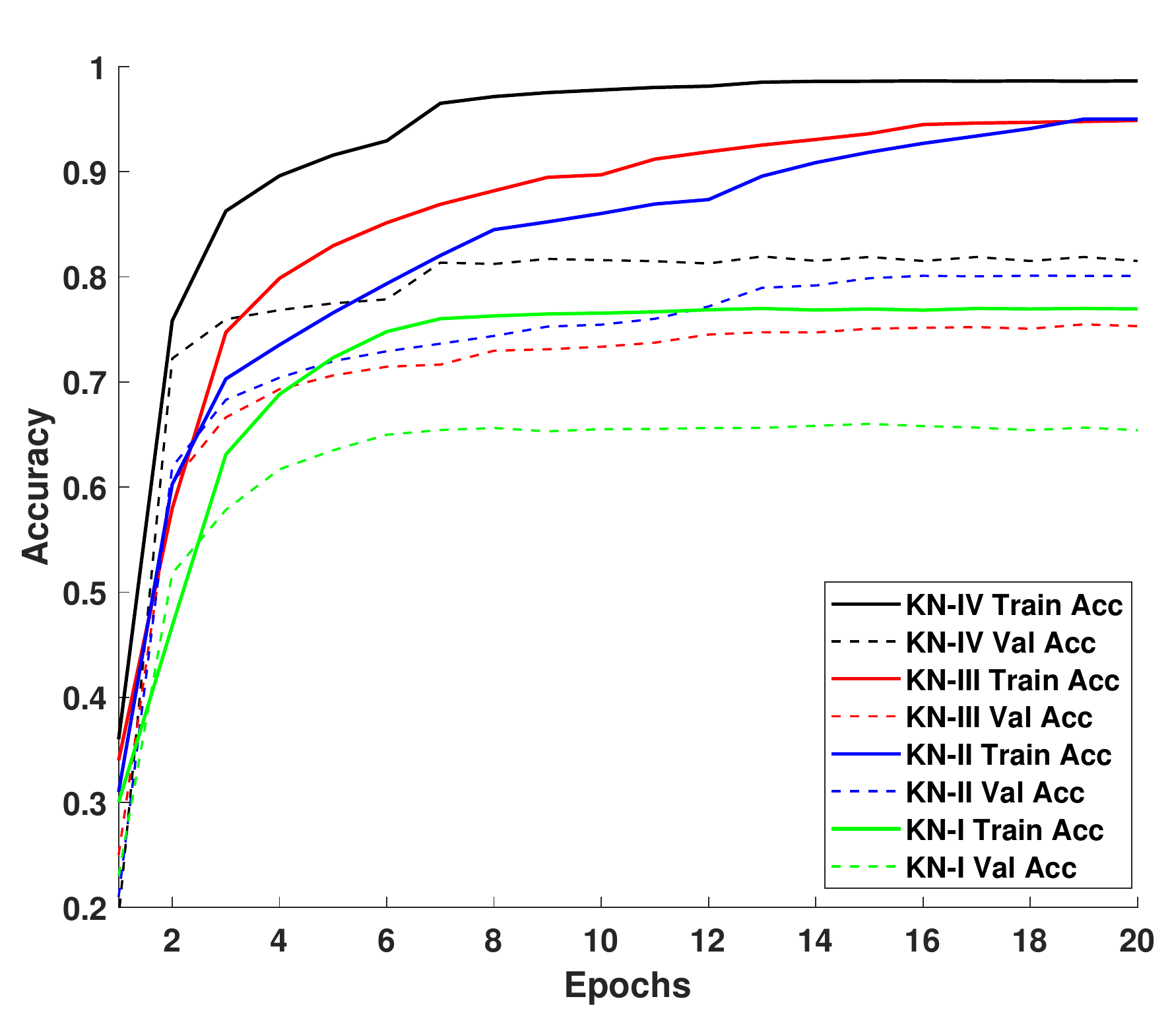}}
\caption{Training/validation accuracy for different KimiaNet configurations.}
\label{fig: covergence}
\end{figure}

Figure \ref{fig: covergence} shows the convergence behaviour of all four models. In most experiments, convergence was observable after 10 epochs. Clearly, the highest accuracy values were achieved when we trained KimiaNet-IV by re-training all DenseNet-121 weights. The general trend of getting higher accuracy by fine-tuning more blocks is also visible. 

%% file: Sections/Experiments.tex
\section{Experiments}
Once the training of different versions of KimiaNet were completed, we used the three datasets to measure the generalization capability of KimiaNet's features extracted from its last pooling layer. The DenseNet features from the same layer were extracted as well for comparison. When searching through features/barcodes to find matches, we used $k$-NN algorithm (with $k=3$) to find the top $k$ matched  (most similar) features/barcodes. The top $k$ matched images through search can be retrieved along with their corresponding metadata. However, we treat the search like a classifier to quantify its performance.

So far, we have mentioned major parameters for KimiaNet. Clearly, any solution based deep networks has many parameters and hyperparameters that need to be adjusted \citep{cui2019new}. In the case of histopathology gigapixel images, additional parameters may be added for operation such as segmentation, patching and clustering \citep{kalra2019yottixel}.

\subsection{TCGA Experiments: Classification through Search}
To evaluate the distinctive power of the provided features, two types of experiments were performed on the testing images: 1) \emph{Horizontal search} which means measuring how accurate the algorithm can find the tomour type across the entire test dataset, and 2) \emph{Vertical search} which is defined as finding the right primary diagnosis of a tomour type among the slides of a specific primary site (containing different primary diagnoses). The features were ``barcoded'' for faster search \citep{tizhoosh2016minmax,kalra2019yottixel}. Barcoding refers to binarization of deep features based on their point-to-point changes (increase/decrease is encoded as 1/0; $a$-$b$ is 1 whereas $b$-$a$ is 0 when $a<b$). Not only binary operations are much faster than arithmetic operations on real-valued features, but also we have already observed that encoding the gradient of deep feature may even increase the matching accuracy \citep{8489574}.

We used the $k$-nearest neighbors algorithm ($k$-NN) approach with Hamming distance to compare the barcoded features of individual patches. For horizontal search the classification accuracy was used whereas for vertical search we calculated the F1 scores. The results for horizontal search are reported in Table \ref{tab:Horizontal} and Figure \ref{fig:Horizontal}.

\input{Tables/TCGA_Table_Horizon_top3}
\vspace{0.12in}
\textbf{Analysis of Horizontal Search --} As Table \ref{tab:Horizontal} shows KimiaNet improves the search accuracy in all cases (on average $41\% \pm 14\%$). A Kolmogorov-Smirnov test of normality delivered a test statistic of $D=0.25079$ for DenseNet ($p=0.27417$) and $D=0.28253$ for KimiaNet IV ($p=0.2433$). Hence, both distributions did not differ significantly from a normal distribution. DN showed an average accuracy of 44.8\% (std=19.9\%) whereas KN-IV delivered an average accuracy of 85.4\% (std=11.6\%). With 68\% improvement, melanocyctic malignancies benefited the most from KimiaNet features. However, the performance for head and neck (63\%) and mesenchymal (54\%) has also shown substantial increase over the average improvement. Although brain has the lowest improvement (27\%), its search accuracy reaches 99\% with KimiaNet compared to 72\% with DenseNet.

\begin{figure}[htb]
\centerline{\includegraphics[width=\columnwidth]{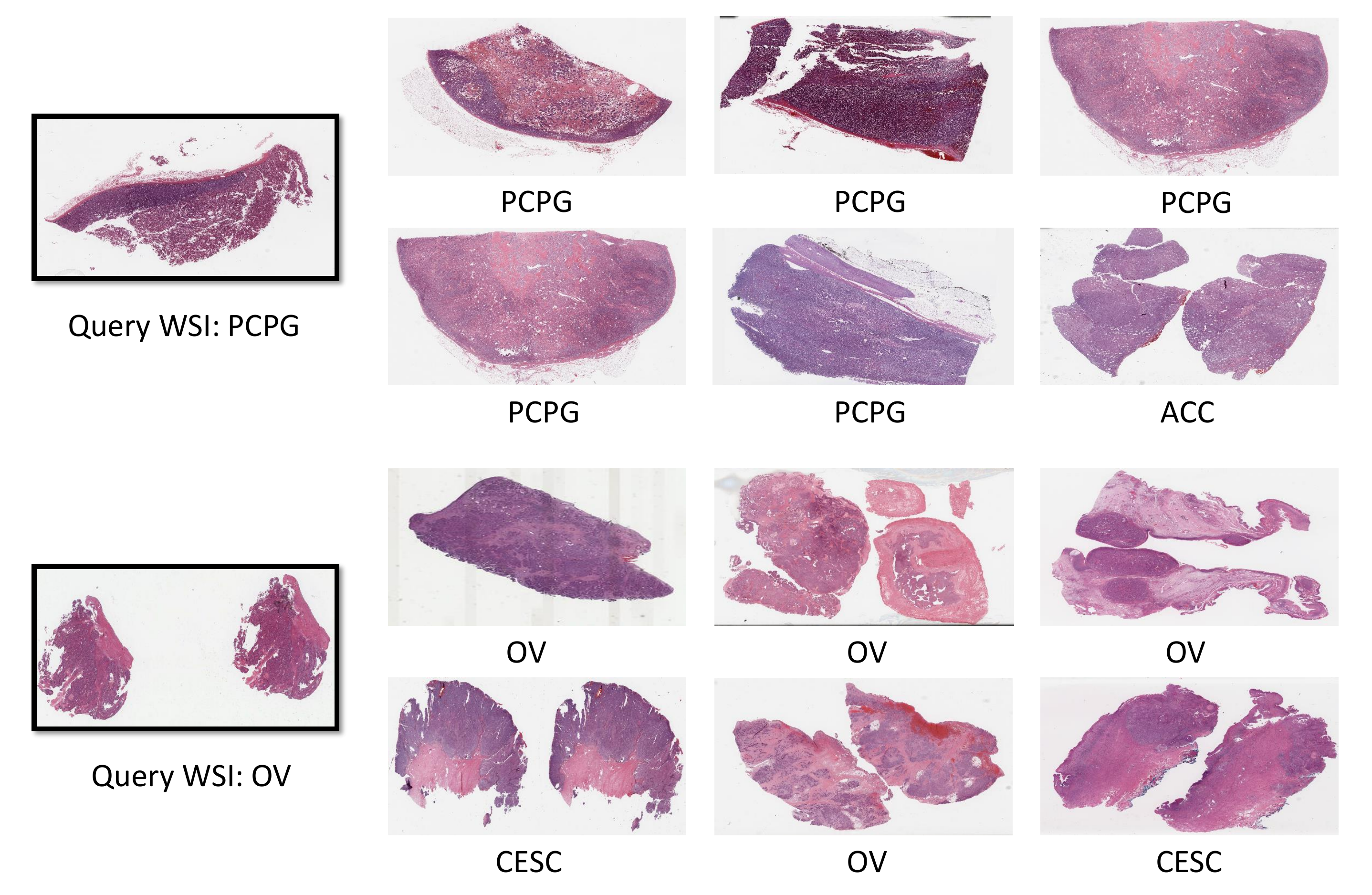}}
\caption{Results for two sample query WSIs (left): Corresponding search results based on KimiaNet features (top row for each query WSI) and DenseNet features (bottom row for each query WSI) and their assigned TCGA primary diagnosis. For TCGA project IDs see Table \ref{tab:diagnosis-abbrv} in Appendix.}
\label{fig:sampleWSIsearch}
\end{figure}

\begin{figure*}[htb]
\centering
\includegraphics[width=0.49\textwidth]{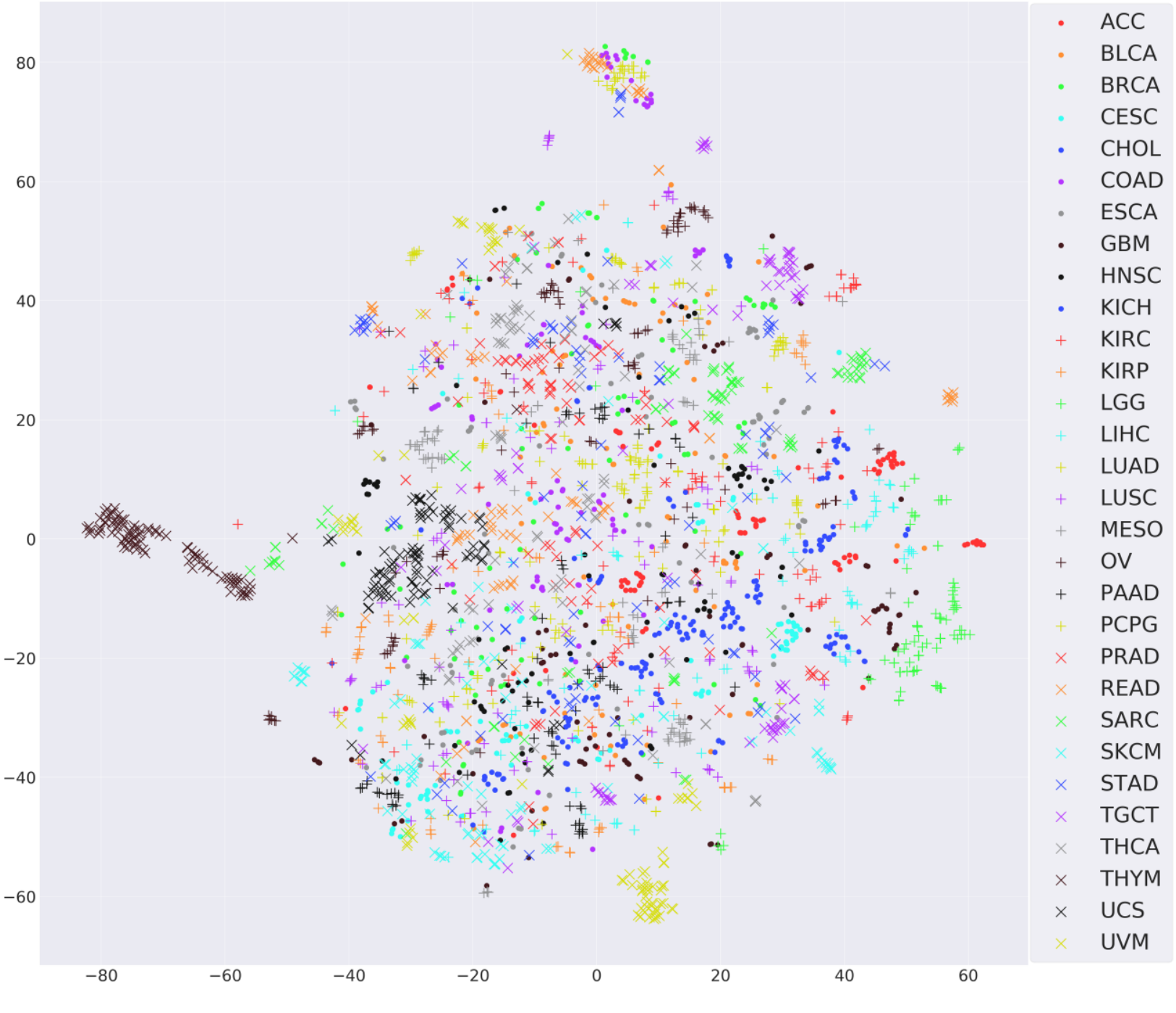}
\includegraphics[width=0.49\textwidth]{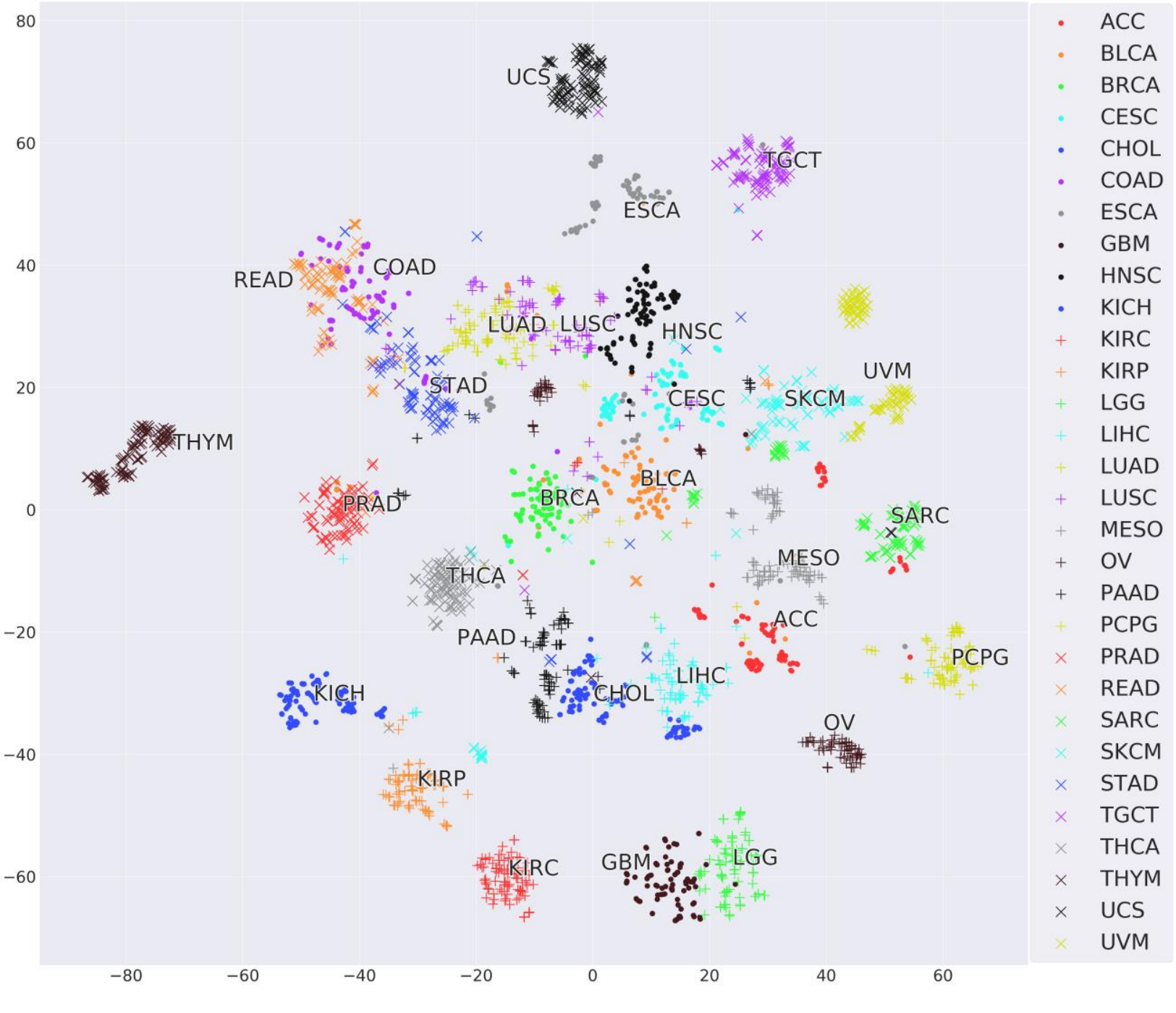}
\caption{t-SNE visualization of randomly selected test images for DenseNet (left) and KimiaNet (right). }
\label{fig:tSNE}
\end{figure*}

\vspace{0.12in}
For cancer subtyping, we run the vertical search where we confined the search to each tumour site to extract the correct diagnosis for each primary site. WSIs were recognized through the ``\emph{median-of-min}'' approach: the minimum Hamming distance for each patch of the query mosaic was calculated when compared to all patches of other WSIs. The median value of all minimum distances was taken as the matching score for the query WSI. Figure \ref{fig:sampleWSIsearch} shows sample queries and results for both DenseNet and KimiaNet. Examination of t-SNE visualization, applied on test images, showed that KimiaNet features have superior class discrimination (Figure \ref{fig:tSNE}). Detailed results are reported in Table \ref{tab:verticalSearch}. Here, we used the F1-measure to account for sensitivity and specificity.   

\input{Tables/TCGA_tables_Vertical_F1}

\textbf{Analysis of Vertical Search --} The higher discrimination power of KimiaNet features is clear. This is not only manifested in the t-SNE visualization (Figure \ref{fig:tSNE}) but also quantified in F1-measures (Table \ref{tab:verticalSearch}). For all subtypes, the F1 score of KimiaNet is higher than DenseNet.

\subsection{Endometrium Data Experiments: Classification}
\begin{figure}[htb]
\centering
\includegraphics[width=1\columnwidth]{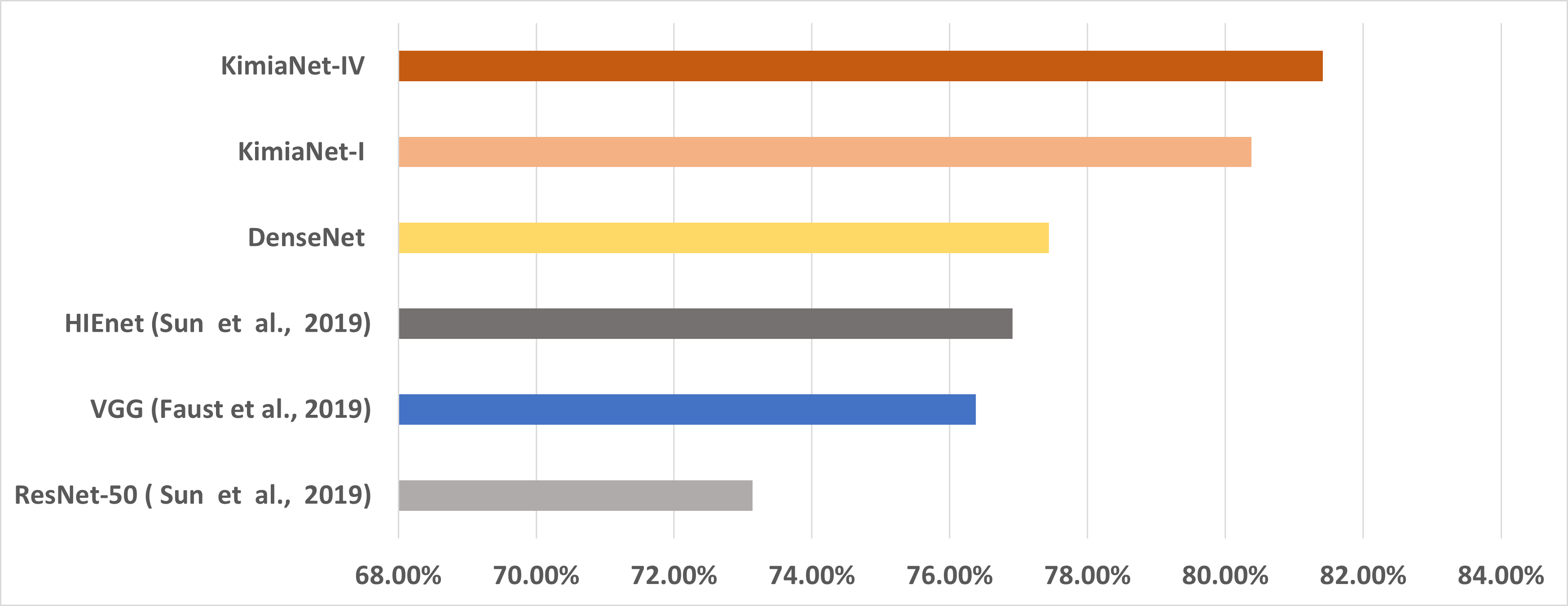}
\caption{Endometrium dataset: SVM accuracy for different deep features: Fine-tuned VGG} vs. DenseNet vs. HIEnet vs. KimiaNet for different input sizes.
\label{fig:endoresults}
\end{figure}

\begin{figure}[htb]
\includegraphics[width=0.50 \columnwidth]{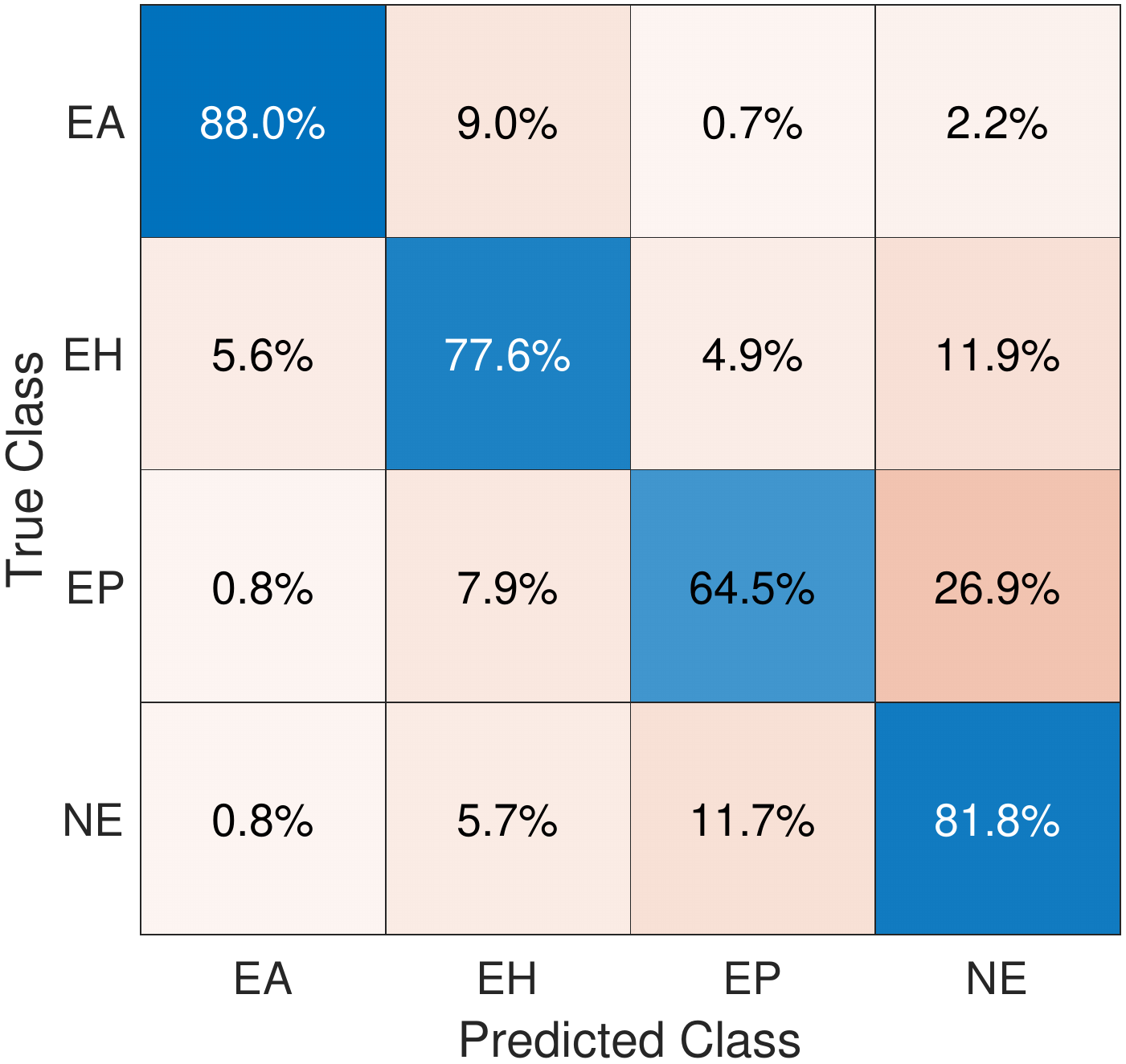}
\includegraphics[width=0.48\columnwidth]{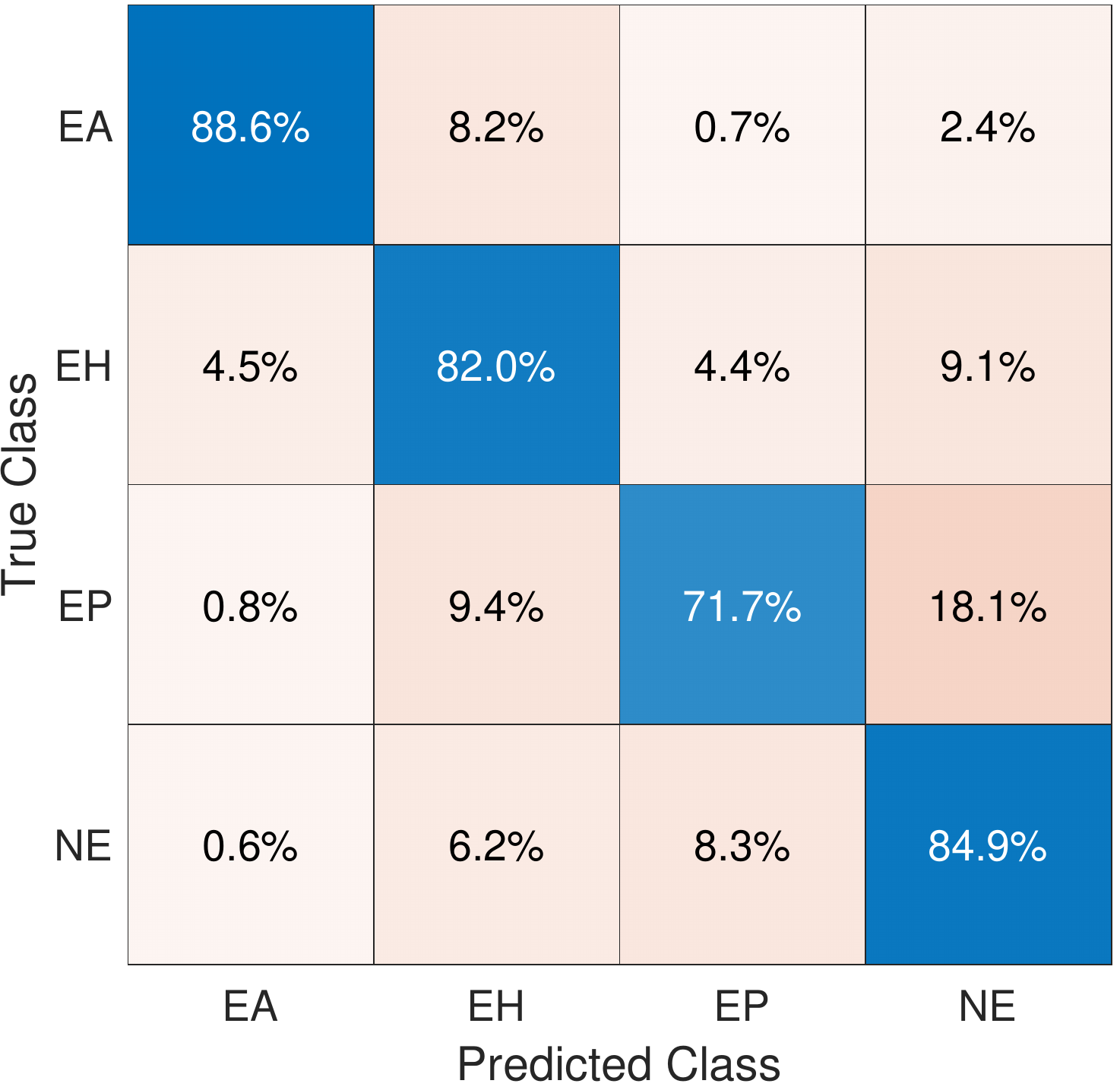}
\caption{Confusion matrices for DenseNet (left) and KimiaNet features (right) for the endometrium dataset.}
\label{fig:ConfEndo}
\end{figure}

To verify the performance of deep features of KimiaNet on the pathology datasets in comparison with other networks, we conducted several experiments on Endometrium images as one of the most recently released datasets. One of these experiments is comparing KimiaNet to another histopathology feature extractor, which is a fine-tuned VGG-19 using 838,644 human-annotated histopathologic patches spanning 74 different lesional and non-lesional tissue types \citep{faust2019intelligent}. We extracted deep features from all patches with the network input default size ($224\!\times\!224$ pixels for DenseNet, $1000\times 1000$ pixels for KimiaNet, and $1024\times 1024$ pixels for the fine-tuned VGG-19). Subsequently, features of each network were used to classify the images using SVM (Support Vector Machines). Ten fold cross-validation with fixed folds, was performed to train the best classifier in each case. The Cubic SVM outperformed all other SVM versions. As Figure \ref{fig:endoresults} demonstrates, with 81.41\% KimiaNet features showed improvement surpassing HIEnet with 76.91\% and the fine-tuned VGG-19 with 76.38\%. 
 Figure \ref{fig:ConfEndo} illustrates the confusion matrices of DenseNet and KimiaNet.

\subsection{Colorectal Data Experiments: Classification}
\begin{figure}[t]
\centerline{\includegraphics[width=0.65\columnwidth]{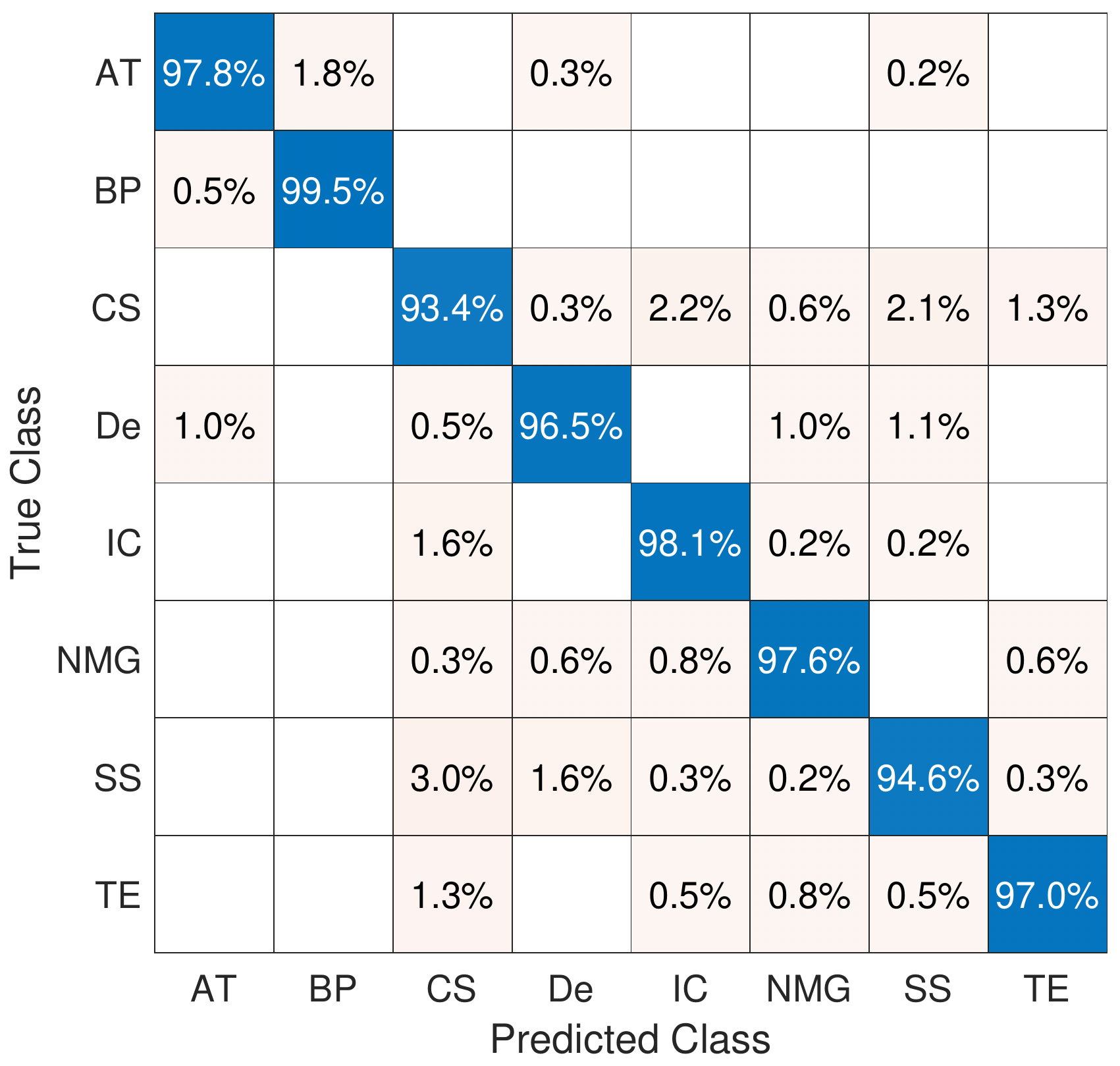}}
\caption{CRC dataset confusion matrix for KimiaNet features.}
\label{fig:CRCconfusion}
\end{figure}

In the last experiment, we repeated the previous experiments with the colorectal dataset. KimiaNet delivered higher accuracy than DenseNet and the fine-tuned VGG-19 (Table \ref{tab:CRCresults}) whereas only an ensemble CNN approach provided higher accuracy. The confusion matrix for KimiaNet shows a pronounced diagonal (Figure \ref{fig:CRCconfusion}). 

\begin{table}[t]
\small
\centering
\caption{Results for colorectal cancer dataset.}
\begin{tabular}{l||c}
\multicolumn{1}{c ||}{Methods} &  Accuracy \\ \hline
Combined features \citep{kather2016multi} & 87.40\% \\
Fine-tuned VGG-19 on 74 classes \citep{faust2019intelligent} & 93.58\% \\
DenseNet  & 94.90\% \\
KN-I & 96.38\% \\
KN-IV & 96.80\% \\
Ensemble of CNNs Here1 \citep{nanni2018ensemble} & 97.60\% 
\end{tabular}
\label{tab:CRCresults}
\end{table}

\subsection{High-Cellularity Mosaic}
The cellMosaic is  a novel approach using weak/soft labels on valuable large datasets like the TCGA that are not annotated at the pixel level. Using the cellMosaic has apparently enabled KimiaNet to outperform DenseNet for TCGA diagnostic images. As we envision using KimiaNet as a feature extractor for histopathology (and not as a classifier), the cellMosaic may also have induced a bias toward high-grade carcinomas and perhaps inflammations with high-cellularity. The additional experiments with two other dataset provided confidence that KimiaNet can in fact represent histopathology patterns in a variety of single patches, and not in cellMosaic as for TCGA whole-slide images. For instance, the CRC dataset contains classes like adipose, debris, normal and background with low to no cellularity at all, and KimiaNet features were able to distinguish them from tumour epithelium with high-cellularity. However, more experiments  with other datasets would be beneficial to more closely define the KimiaNet’s application domain.

\input{Tables/Table_CBR_Nets}

\subsection{CBR Nets: Small versus large}

According to recently published results, smaller topologies may in fact provide better or the same results for medical images as delivered by large networks pre-trained and tested using datasets like ImageNet \citep{raghu2019transfusion}.

Compact and simple networks made of repetitions of convolutional, batch-normalization and ReLU layers, also called CBR nets, have been found to be quite accurate, compared to larger standard ImageNet models, for applications such as retinal and x-ray image identifications. \cite{raghu2019transfusion} investigated large networks like ResNet50 (\raisebox{-0.7ex}{\textasciitilde}23.5M weights) and Inception-V3 (\raisebox{-0.7ex}{\textasciitilde}23M weights) and showed that, for instance, the Small CBR network (\raisebox{-0.7ex}{\textasciitilde}2M weights) generated comparable results. The authors, however, did not investigate the popular DenseNet121 (\raisebox{-0.7ex}{\textasciitilde}7M weights). In order to verify our results in light of these recent experiments, we did implement and test multiple CBR network topologies to validate against KimiaNet.

We implemented the Small, LargeT and LargeW topologies as suggested by \cite{raghu2019transfusion} with a slight modification of adding a dense layer before the last fully connected layer\footnote{The performance of CBR features was very low without the additional dense layer.}. We also implemented a modified Small network by adding the fifth CBR block to the Small topology. The networks were initialized with random weights and trained from scratch for around 40 epochs using the same TCGA dataset as used for KimiaNet. All settings were the same as for the training of KimiaNet except that the learning rate was initialized with 0.003 and the batch size was 32.

Tables \ref{tab:CBR_Nets_Horizontal} shows the results of the experiments for horizontal and vertical search. For horizontal search, it can be observed that KimiaNet achieves a considerably higher accuracy average with more consistency than all other networks, having a 30\% difference with the second best results (LargeT with around 1.5 M parameters more than KimiaNet). In addition, KimiaNet achieves the maximum accuracy for all of the classes (12 in horizontal search) while none of other networks reached the maximum accuracy for any of the classes. There is a similar pattern for vertical search. The average of F1 scores for KimiaNet is 10\% higher than the second best network, LargeW, which, again, has more parameters than KimiaNet. KimiaNet's low standard deviation of F1 scores shows its stability compared to other networks and the number of maximum F1 scores (23 out of 26 classes) confirms its superiority over CBR networks.

%% file: Tables/TCGA_Table_Horizon_top3.tex
\begin{table}[t]
\centering
\scriptsize
\caption{3-Nearest Neighbors  accuracy (\%) for the horizontal search among 744 WSIs for differently fine-tuned/trained KimiaNet. The best results are highlighted. The last column shows the improvement of accuracy (\%) through KimiaNet compared to DenseNet.}

\begin{tabular}{lc||c||cccc||c}
Tumor Type       & Patient \# & DN                                 & I                                 & II                                 & III                                 & IV                                 & diff  \\ 
\hline
Brain            & 74         & 72 & 96                                 & 97                                 & {\cellcolor[rgb]{0.749,1,0.749}}99 & {\cellcolor[rgb]{0.749,1,0.749}} 99                                 & +27   \\
Breast           & 91         & 53 & 86                                 & 87                                 & {\cellcolor[rgb]{0.749,1,0.749}}91 & {\cellcolor[rgb]{0.749,1,0.749}} 91                                 & +38   \\
Endocrine        & 72         & 65 & 86                                 & 89                                 & {\cellcolor[rgb]{0.749,1,0.749}}93 & 92                                 & +28   \\
Gastro. & 88         & 53                                 & 74                                 & 81                                 & 80                                 & {\cellcolor[rgb]{0.749,1,0.749}}84 & +31   \\
Gynaec.   & 30         & 13 & 43                                 & 40                                 & 47                                 & {\cellcolor[rgb]{0.749,1,0.749}}57 & +44   \\
Head/neck    & 32         & 25 & 75                                 & 69                                 & 81                                 & {\cellcolor[rgb]{0.749,1,0.749}}88 & +63   \\
Liver            & 51         & 43 & 67                                 & 69                                 & 80                                 & {\cellcolor[rgb]{0.749,1,0.749}}88 & +45   \\
Melanocytic      & 28         & 18 & 57                                 & 54                                 & 75                                 & {\cellcolor[rgb]{0.749,1,0.749}}86 & +68   \\
Mesenchymal      & 13         & 23 & 38                                      & 62                                 & {\cellcolor[rgb]{0.749,1,0.749}} 69              &    {\cellcolor[rgb]{0.749,1,0.749}}    69           & +46   \\
Prostate/testis  & 53         & 57 & 89                                 & 91                                 & 94                                 & {\cellcolor[rgb]{0.749,1,0.749}}96 & +39   \\
Pulmonary        & 86         & 56 & 83                                 & {\cellcolor[rgb]{0.749,1,0.749}}86 & 85                                 & {\cellcolor[rgb]{0.749,1,0.749}}86                                 & +30   \\
Urinary tract    & 123        & 59 & {\cellcolor[rgb]{0.749,1,0.749}}89 & {\cellcolor[rgb]{0.749,1,0.749}}89                                 & 88                                 & {\cellcolor[rgb]{0.749,1,0.749}}89                                 & +30  
\end{tabular}
\label{tab:Horizontal}
\end{table} 

\begin{figure}[t]
\centerline{\includegraphics[width=\columnwidth]{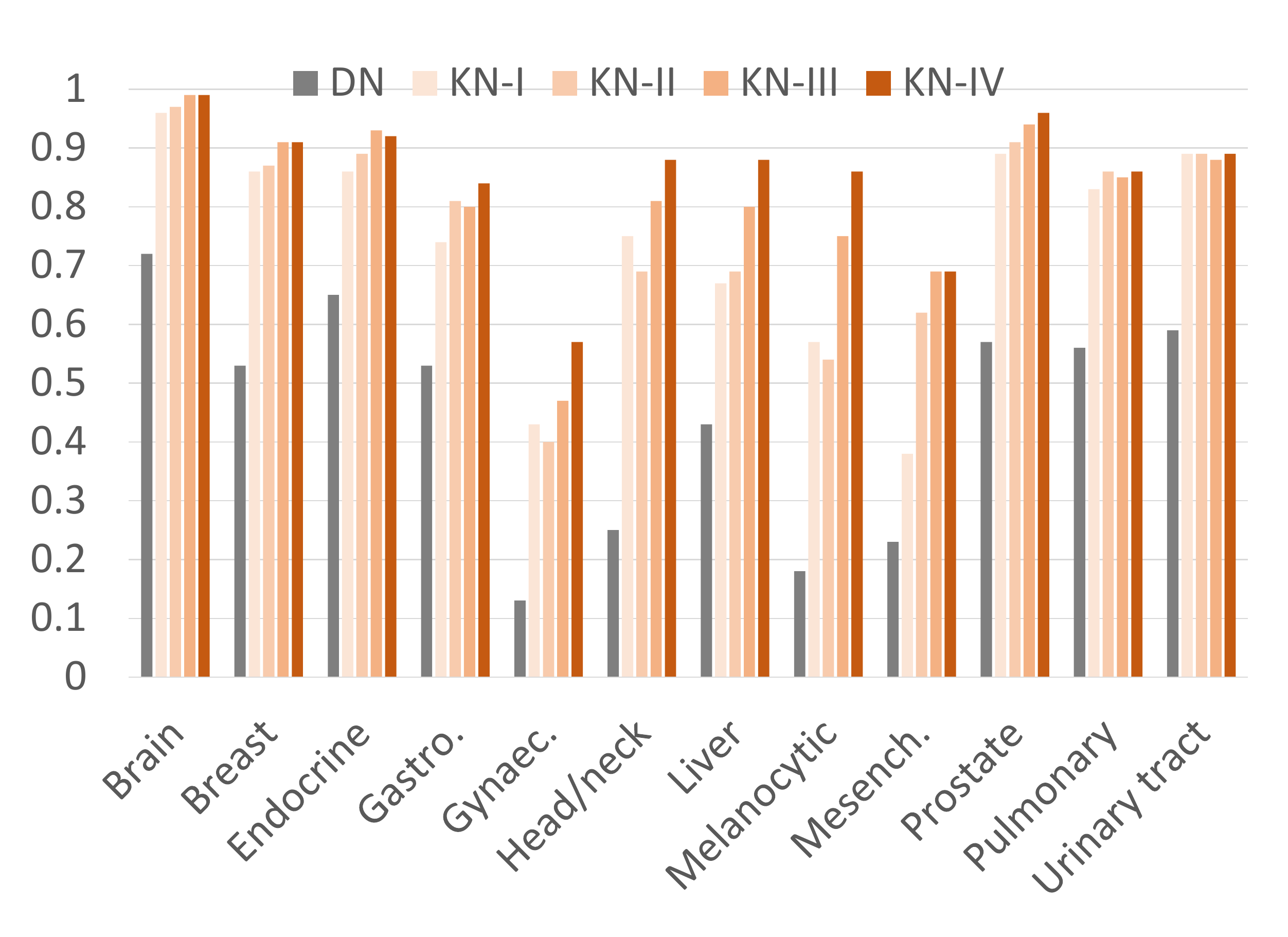}}
\caption{Horizontal search results (accuracy, in percentage) for TCGA data.}
\label{fig:Horizontal}
\end{figure}

%% file: Tables/TCGA_tables_Vertical_F1.tex
\begin{table}[htb]
{\scriptsize
\centering

\caption{$k$-NN results for the vertical search among 744 WSIs. The best results are highlighted. F1-measure has been reported here instead of simple classification accuracy. For TCGA codes see Table \ref{tab:diagnosis-abbrv} in Appendix.}
\label{tab:verticalSearch}
\begin{tabular}{llr|c|rrrr}
Site & Subtype & $n_\textrm{slides}$ & DN & I & II & III & IV \\ 
\hline 
Brain                     & LGG  & 39                    &   71                          &   75                     &   82                      & {\cellcolor[rgb]{.749,1,  .749}}  85    &   81                       \\
Brain                     & GBM  & 35                    &   77                          &   73                     &   80                       & {\cellcolor[rgb]{.749,1,  .749}}  83    &   81                       \\ 
\hline
Endocrine                 & THCA & 51                    &   94                          &   98                     &   98                      &   99                       & {\cellcolor[rgb]{.749,1,  .749}}100       \\
Endocrine                 & ACC  & 6                     &   25                          &   25                     &   20                       & {\cellcolor[rgb]{.749,1,  .749}}  55    &   44                       \\
Endocrine                 & PCPG & 15                    &   57                          &   75                     &   73                      &   80                        & {\cellcolor[rgb]{.749,1,  .749}}  85    \\ 
\hline
Gastro.    & ESCA & 14                    &   50                           &   73                     &   50                       & {\cellcolor[rgb]{.749,1,  .749}}  83    &   78                       \\
Gastro.    & COAD & 32                    &   65                          & {\cellcolor[rgb]{.749,1,  .749}}  76  &   75                      &   75                       & {\cellcolor[rgb]{.749,1,  .749}}  76    \\ 

Gastro.    & STAD & 30                    &   63                          &   77                     &   73                      &   84                       & {\cellcolor[rgb]{.749,1,  .749}}  86    \\
Gastro.    & READ & 12                    &   22                          & {\cellcolor[rgb]{.749,1,  .749}}  30  &   26                      &   29                       & {\cellcolor[rgb]{.749,1,  .749}}  30     \\ 
\hline
Gynaeco.            & UCS  & 3                     &   75                          & {\cellcolor[rgb]{.749,1,  .749}}  86  &   60                       &   75                       & {\cellcolor[rgb]{.749,1,  .749}}  86    \\
Gynaeco.            & CESC & 17                    &   88                          & {\cellcolor[rgb]{.749,1,  .749}}  97  &   84                      &   97                       &   94                       \\
Gynaeco.            & OV   & 10                    &   67                          &   89                     &   74                      & {\cellcolor[rgb]{.749,1,  .749}}  95    &   {\cellcolor[rgb]{.749,1,  .749}}95                       \\ 
\hline
Liver, panc. & CHOL & 4                     &   29                          & {\cellcolor[rgb]{.749,1,  .749}}  40  &   31                      & {\cellcolor[rgb]{.749,1,  .749}}  40     & {\cellcolor[rgb]{.749,1,  .749}}  40     \\
Liver, panc. & LIHC & 35                    &   86                          &   94                     &   87                      & {\cellcolor[rgb]{.749,1,  .749}}  97    &   96                       \\
Liver, panc. & PAAD & 12                    &   70                           &   73                     &   56                      & {\cellcolor[rgb]{.749,1,  .749}}  82    &   76                       \\ 
\hline
Melanocytic   & SKCM & 24                    &   92                          &   94                     &   94                      & {\cellcolor[rgb]{.749,1,  .749}}  98    &   94                       \\
Melanocytic  & UVM  & 4                     & 0                             &   40                     &   40                       & {\cellcolor[rgb]{.749,1,  .749}}  86    &   67                       \\ 
\hline
Prostate/testis           & PRAD & 40                    &   99                          & {\cellcolor[rgb]{.749,1,  .749}}100  &   99                      & {\cellcolor[rgb]{.749,1,  .749}}100       & {\cellcolor[rgb]{.749,1,  .749}}100       \\
Prostate/testis           & TGCT & 13                    &   96                          & {\cellcolor[rgb]{.749,1,  .749}}100  &   96                      & {\cellcolor[rgb]{.749,1,  .749}}100       & {\cellcolor[rgb]{.749,1,  .749}}100       \\ 
\hline
Pulmonary                 & LUAD & 38                    &   65                          &   73                     &   72                      &   69                       & {\cellcolor[rgb]{.749,1,  .749}}  78    \\
Pulmonary                 & LUSC & 43                    &   69                          &   74                     &   74                      &   75                       & {\cellcolor[rgb]{.749,1,  .749}}  84    \\
Pulmonary                 & MESO & 5                     & 0                             &   0                     & 0                         &   33                       & {\cellcolor[rgb]{.749,1,  .749}}  75    \\ 
\hline
Urinary tract             & BLCA & 34                    &   90                           & {\cellcolor[rgb]{.749,1,  .749}}  96  &   93                      &   93                       &   93                       \\
Urinary tract             & KIRC & 50                    &   83                          &   95                     & {\cellcolor[rgb]{.749,1,  .749}}  99   &   97                       &   97                       \\
Urinary tract             & KIRP & 28                    &   77                          & {\cellcolor[rgb]{.749,1,  .749}}  91  & {\cellcolor[rgb]{.749,1,  .749}}  91   & {\cellcolor[rgb]{.749,1,  .749}}  91    & {\cellcolor[rgb]{.749,1,  .749}}  91    \\
Urinary tract             & KICH & 11                    &   48                          & {\cellcolor[rgb]{.749,1,  .749}}  86  &   78                      &   84                       & {\cellcolor[rgb]{.749,1,  .749}}  86    \\ 
\hline

\end{tabular}
}

\end{table}

%% file: Tables/Table_CBR_Nets.tex
\begin{table}[t]
\centering
\scriptsize
\caption{Average and standard deviation of achieved score (accuracy for horizontal and F1 score for vertical search) for each topology for TCGA images. $n_\textrm{max}$ is number of times maximum score achieved (out of 12 and 26 classes for horizontal and vertical search respectively) and $n_\textrm{weights}$ is the approximate number of parameters for the topology.}
\resizebox{0.99 \columnwidth}{!}{
\begin{tabular}{l || c  c || c c || c}
Topology & Horizontal & $n_\textrm{max}$ & Vertical & $n_\textrm{max}$ & $n_\textrm{weights}$ \\ \hline
CBR Small          & 48$\pm$20 & 0  & 69$\pm$24 & 1  & 2   Millions  \\
CBR Mod. Small     & 46$\pm$22 & 0  & 68$\pm$29 & 1  & 5.5 Millions \\ 
CBR LargeT         & 55$\pm$20 & 0  & 69$\pm$23 & 3  & 8.5 Millions \\ 
CBR LargeW         & 50$\pm$22 & 0  & 71$\pm$24 & 2  & 8.5 Millions \\ \hline
DN                 & 45$\pm$20 & 0  & 64$\pm$28 & 0  & 7   Millions \\
KN IV              & 85$\pm$12 & 12 & 81$\pm$18 & 23 & 7   Millions
\end{tabular}
}
\label{tab:CBR_Nets_Horizontal}
\end{table} 

%% file: Sections/Summary_and_Conclusions.tex
\section{Summary and Conclusions}
The question of image representation is generally an important topic in computer vision and becomes critical in digital pathology due to the texture complexity, polymorphism, and the sheer size of WSIs. Examining recent works clearly shows that high-level embeddings in artificial neural networks are considered the most robust and expressive source for image representation. Pre-trained networks such as DenseNet that draw their discrimination power from intensive training with millions of natural (non-medical) images have found widespread usage in medical image analysis. Several attempts have been reported in literature to fine-tune or train deep networks with histopathology images, a desirable task that is impeded by lack of labelled image data and the need for high-performance computing devices.    

In this work, we proposed KimiaNet in several fine-tuned configurations by using a clustering-based mosaic structure for image representation modified by relaying on high cellularity in order to enable the usage of WSI-level labels in archives with no pixel-level annotations. Three public datasets were employed to generate the results. While there are several approaches to the application of neural networks to medical image analysis, the narrow focus on a single problem/approach within a given study limits guidance on how best to optimize neural networks for pathology tasks. 

Our main contribution in this paper was exploiting a diverse, multi-organ public image repository like TCGA at 20$\times$ magnification to extract large patches, 1000 $\times$ 1000 pixels at high resolution, for training a densely connected network with weak labels to serve as feature extractor. As well, we showed that fine-tuning a deep network on a  sufficiently large number of histopathology images delivers better performance than a pre-trained network model. Finally, we think the publicly available data and code are valuable for the computational pathology community. The proposed high-cellularity mosaic crucially facilitated the training but may have introduced a bias toward certain histologic features at the expense of visual similarities. Future works have to carve out a comprehensive list of histopathology applications that could benefit from KimiaNet for image representation.   

\vspace{0.1in}
\textbf{Acknowledgements --} The authors would like to thank the Ontario government for the ORF-RE (Ontario Research Fund - Research Excellence) and NSERC (Natural Sciences and Engineering Research Council of Canada) that have funded this research. 

\vspace{0.1in}
\textbf{Data/Code Availability --} KimiaNet and the folds for the validation of all  datasets will be available for download from ``\emph{https://kimia.uwaterloo.ca}''.

%% file: Tables/tab-diagnosis-abbrv.tex
\begin{table}[h]
\centering
\scriptsize

\resizebox{0.9\columnwidth}{!}{

\begin{tabular}{lp{3cm}c}

Code &                                  Primary Diagnosis &  \#Patients \\
\hline
ACC                      &                           Adrenocortical Carcinoma &       86 \\
BLCA                     &                       Bladder Urothelial Carcinoma &      410 \\
BRCA                     &                          Breast Invasive Carcinoma &     1097 \\
CESC                     &  Cervical Squamous Cell Carcinoma and Endocervical Adenoc.  &      304 \\
CHOL                     &                                 Cholangiocarcinoma &       51 \\
COAD                     &                               Colon Adenocarcinoma &      459 \\
DLBC                     &    Lymphoid Neoplasm Diffuse Large B-cell Lymphoma &       48 \\
ESCA                     &                               Esophageal Carcinoma &      185 \\
GBM                      &                            Glioblastoma Multiforme &      604 \\
HNSC                     &              Head and Neck Squamous Cell Carcinoma &      473 \\
KICH                     &                                 Kidney Chromophobe &      112 \\
KIRC                     &                  Kidney Renal Clear Cell Carcinoma &      537 \\
KIRP                     &              Kidney Renal Papillary Cell Carcinoma &      290 \\
LGG                      &                           Brain Lower Grade Glioma &      513 \\
LIHC                     &                     Liver Hepatocellular Carcinoma &      376 \\
LUAD                     &                                Lung Adenocarcinoma &      522 \\
LUSC                     &                       Lung Squamous Cell Carcinoma &      504 \\
MESO                     &                                       Mesothelioma &       86 \\
OV                       &                  Ovarian Serous Cystadenocarcinoma &      590 \\
PAAD                     &                          Pancreatic Adenocarcinoma &      185 \\
PCPG                     &                 Pheochromocytoma and Paraganglioma &      179 \\
PRAD                     &                            Prostate Adenocarcinoma &      499 \\
READ                     &                              Rectum Adenocarcinoma &      170 \\
SARC                     &                                            Sarcoma &      261 \\
SKCM                     &                            Skin Cutaneous Melanoma &      469 \\
STAD                     &                             Stomach Adenocarcinoma &      442 \\
TGCT                     &                        Testicular Germ Cell Tumors &      150 \\
THCA                     &                                  Thyroid Carcinoma &      507 \\
THYM                     &                                            Thymoma &      124 \\
UCEC                     &               Uterine Corpus Endometrial Carcinoma &      558 \\
UCS                      &                             Uterine Carcinosarcoma &       57 \\
UVM                      &                                     Uveal Melanoma &       80 \\
\bottomrule
\end{tabular}
}

\parbox{0.9\columnwidth}{
\caption{The TCGA codes (in alphabetical order) of all 32 primary diagnoses and corresponding number of evidently diagnosed patients in the dataset (TCGA = The Cancer Genome Atlas)}} 
 \label{tab:diagnosis-abbrv}
\end{table}

%% file: kimianet.bbl
\begin{thebibliography}{54}
\expandafter\ifx\csname natexlab\endcsname\relax\def\natexlab#1{#1}\fi
\providecommand{\url}[1]{\texttt{#1}}
\providecommand{\href}[2]{#2}
\providecommand{\path}[1]{#1}
\providecommand{\DOIprefix}{doi:}
\providecommand{\ArXivprefix}{arXiv:}
\providecommand{\URLprefix}{URL: }
\providecommand{\Pubmedprefix}{pmid:}
\providecommand{\doi}[1]{\href{http://dx.doi.org/#1}{\path{#1}}}
\providecommand{\Pubmed}[1]{\href{pmid:#1}{\path{#1}}}
\providecommand{\bibinfo}[2]{#2}
\ifx\xfnm\relax \def\xfnm[#1]{\unskip,\space#1}\fi
\bibitem[{Almagro~Armenteros et~al.(2017)Almagro~Armenteros, S{\o}nderby,
  S{\o}nderby, Nielsen and Winther}]{almagro2017deeploc}
\bibinfo{author}{Almagro~Armenteros, J.J.}, \bibinfo{author}{S{\o}nderby,
  C.K.}, \bibinfo{author}{S{\o}nderby, S.K.}, \bibinfo{author}{Nielsen, H.},
  \bibinfo{author}{Winther, O.}, \bibinfo{year}{2017}.
\newblock \bibinfo{title}{Deeploc: prediction of protein subcellular
  localization using deep learning}.
\newblock \bibinfo{journal}{Bioinformatics} \bibinfo{volume}{33},
  \bibinfo{pages}{3387--3395}.
\bibitem[{Babaie and Tizhoosh(2019)}]{babaie2019deep}
\bibinfo{author}{Babaie, M.}, \bibinfo{author}{Tizhoosh, H.R.},
  \bibinfo{year}{2019}.
\newblock \bibinfo{title}{Deep features for tissue-fold detection in
  histopathology images}.
\newblock \bibinfo{journal}{arXiv preprint arXiv:1903.07011} .
\bibitem[{Bilaloglu et~al.(2019)Bilaloglu, Wu, Fierro, Sanchez, Ocampo,
  Razavian, Coudray and Tsirigos}]{bilaloglu2019efficient}
\bibinfo{author}{Bilaloglu, S.}, \bibinfo{author}{Wu, J.},
  \bibinfo{author}{Fierro, E.}, \bibinfo{author}{Sanchez, R.D.},
  \bibinfo{author}{Ocampo, P.S.}, \bibinfo{author}{Razavian, N.},
  \bibinfo{author}{Coudray, N.}, \bibinfo{author}{Tsirigos, A.},
  \bibinfo{year}{2019}.
\newblock \bibinfo{title}{Efficient pan-cancer whole-slide image classification
  and outlier detection using convolutional neural networks}.
\newblock \bibinfo{journal}{bioRxiv} , \bibinfo{pages}{633123}.
\bibitem[{Burt et~al.(2018)Burt, Torosdagli, Khosravan, RaviPrakash, Mortazi,
  Tissavirasingham, Hussein and Bagci}]{burt2018deep}
\bibinfo{author}{Burt, J.R.}, \bibinfo{author}{Torosdagli, N.},
  \bibinfo{author}{Khosravan, N.}, \bibinfo{author}{RaviPrakash, H.},
  \bibinfo{author}{Mortazi, A.}, \bibinfo{author}{Tissavirasingham, F.},
  \bibinfo{author}{Hussein, S.}, \bibinfo{author}{Bagci, U.},
  \bibinfo{year}{2018}.
\newblock \bibinfo{title}{Deep learning beyond cats and dogs: recent advances
  in diagnosing breast cancer with deep neural networks}.
\newblock \bibinfo{journal}{The British journal of radiology}
  \bibinfo{volume}{91}, \bibinfo{pages}{20170545}.
\bibitem[{Campanella et~al.(2019)Campanella, Hanna, Geneslaw, Miraflor, Silva,
  Busam, Brogi, Reuter, Klimstra and Fuchs}]{campanella2019clinical}
\bibinfo{author}{Campanella, G.}, \bibinfo{author}{Hanna, M.G.},
  \bibinfo{author}{Geneslaw, L.}, \bibinfo{author}{Miraflor, A.},
  \bibinfo{author}{Silva, V.W.K.}, \bibinfo{author}{Busam, K.J.},
  \bibinfo{author}{Brogi, E.}, \bibinfo{author}{Reuter, V.E.},
  \bibinfo{author}{Klimstra, D.S.}, \bibinfo{author}{Fuchs, T.J.},
  \bibinfo{year}{2019}.
\newblock \bibinfo{title}{Clinical-grade computational pathology using weakly
  supervised deep learning on whole slide images}.
\newblock \bibinfo{journal}{Nature medicine} \bibinfo{volume}{25},
  \bibinfo{pages}{1301--1309}.
\bibitem[{Cooper et~al.(2018)Cooper, Demicco, Saltz, Powell, Rao and
  Lazar}]{cooper2018pancancer}
\bibinfo{author}{Cooper, L.A.}, \bibinfo{author}{Demicco, E.G.},
  \bibinfo{author}{Saltz, J.H.}, \bibinfo{author}{Powell, R.T.},
  \bibinfo{author}{Rao, A.}, \bibinfo{author}{Lazar, A.J.},
  \bibinfo{year}{2018}.
\newblock \bibinfo{title}{Pancancer insights from the cancer genome atlas: the
  pathologist's perspective}.
\newblock \bibinfo{journal}{The Journal of pathology} \bibinfo{volume}{244},
  \bibinfo{pages}{512--524}.
\bibitem[{Coudray et~al.(2018)Coudray, Ocampo, Sakellaropoulos, Narula,
  Snuderl, Feny{\"o}, Moreira, Razavian and
  Tsirigos}]{coudray2018classification}
\bibinfo{author}{Coudray, N.}, \bibinfo{author}{Ocampo, P.S.},
  \bibinfo{author}{Sakellaropoulos, T.}, \bibinfo{author}{Narula, N.},
  \bibinfo{author}{Snuderl, M.}, \bibinfo{author}{Feny{\"o}, D.},
  \bibinfo{author}{Moreira, A.L.}, \bibinfo{author}{Razavian, N.},
  \bibinfo{author}{Tsirigos, A.}, \bibinfo{year}{2018}.
\newblock \bibinfo{title}{Classification and mutation prediction from
  non--small cell lung cancer histopathology images using deep learning}.
\newblock \bibinfo{journal}{Nature medicine} \bibinfo{volume}{24},
  \bibinfo{pages}{1559}.
\bibitem[{Cui and Bai(2019)}]{cui2019new}
\bibinfo{author}{Cui, H.}, \bibinfo{author}{Bai, J.}, \bibinfo{year}{2019}.
\newblock \bibinfo{title}{A new hyperparameters optimization method for
  convolutional neural networks}.
\newblock \bibinfo{journal}{Pattern Recognition Letters} \bibinfo{volume}{125},
  \bibinfo{pages}{828--834}.
\bibitem[{Deng et~al.(2009)Deng, Dong, Socher, Li, Li and
  Fei-Fei}]{deng2009imagenet}
\bibinfo{author}{Deng, J.}, \bibinfo{author}{Dong, W.},
  \bibinfo{author}{Socher, R.}, \bibinfo{author}{Li, L.J.},
  \bibinfo{author}{Li, K.}, \bibinfo{author}{Fei-Fei, L.},
  \bibinfo{year}{2009}.
\newblock \bibinfo{title}{Imagenet: A large-scale hierarchical image database},
  in: \bibinfo{booktitle}{2009 IEEE conference on computer vision and pattern
  recognition}, \bibinfo{organization}{Ieee}. pp. \bibinfo{pages}{248--255}.
\bibitem[{Donahue et~al.(2014)Donahue, Jia, Vinyals, Hoffman, Zhang, Tzeng and
  Darrell}]{donahue2014decaf}
\bibinfo{author}{Donahue, J.}, \bibinfo{author}{Jia, Y.},
  \bibinfo{author}{Vinyals, O.}, \bibinfo{author}{Hoffman, J.},
  \bibinfo{author}{Zhang, N.}, \bibinfo{author}{Tzeng, E.},
  \bibinfo{author}{Darrell, T.}, \bibinfo{year}{2014}.
\newblock \bibinfo{title}{Decaf: A deep convolutional activation feature for
  generic visual recognition}, in: \bibinfo{booktitle}{International conference
  on machine learning}, pp. \bibinfo{pages}{647--655}.
\bibitem[{Faust et~al.(2019)Faust, Bala, van Ommeren, Portante, Al~Qawahmed,
  Djuric and Diamandis}]{faust2019intelligent}
\bibinfo{author}{Faust, K.}, \bibinfo{author}{Bala, S.}, \bibinfo{author}{van
  Ommeren, R.}, \bibinfo{author}{Portante, A.}, \bibinfo{author}{Al~Qawahmed,
  R.}, \bibinfo{author}{Djuric, U.}, \bibinfo{author}{Diamandis, P.},
  \bibinfo{year}{2019}.
\newblock \bibinfo{title}{Intelligent feature engineering and ontological
  mapping of brain tumour histomorphologies by deep learning}.
\newblock \bibinfo{journal}{Nature Machine Intelligence} \bibinfo{volume}{1},
  \bibinfo{pages}{316--321}.
\bibitem[{Faust et~al.(2018)Faust, Xie, Han, Goyle, Volynskaya, Djuric and
  Diamandis}]{faust2018visualizing}
\bibinfo{author}{Faust, K.}, \bibinfo{author}{Xie, Q.}, \bibinfo{author}{Han,
  D.}, \bibinfo{author}{Goyle, K.}, \bibinfo{author}{Volynskaya, Z.},
  \bibinfo{author}{Djuric, U.}, \bibinfo{author}{Diamandis, P.},
  \bibinfo{year}{2018}.
\newblock \bibinfo{title}{Visualizing histopathologic deep learning
  classification and anomaly detection using nonlinear feature space
  dimensionality reduction}.
\newblock \bibinfo{journal}{BMC bioinformatics} \bibinfo{volume}{19},
  \bibinfo{pages}{173}.
\bibitem[{Fu et~al.(2019)Fu, Jung, Torne, Gonzalez, Vohringer, Jimenez-Linan,
  Moore and Gerstung}]{fu2019pan}
\bibinfo{author}{Fu, Y.}, \bibinfo{author}{Jung, A.W.}, \bibinfo{author}{Torne,
  R.V.}, \bibinfo{author}{Gonzalez, S.}, \bibinfo{author}{Vohringer, H.},
  \bibinfo{author}{Jimenez-Linan, M.}, \bibinfo{author}{Moore, L.},
  \bibinfo{author}{Gerstung, M.}, \bibinfo{year}{2019}.
\newblock \bibinfo{title}{Pan-cancer computational histopathology reveals
  mutations, tumor composition and prognosis}.
\newblock \bibinfo{journal}{bioRxiv} , \bibinfo{pages}{813543}.
\bibitem[{Glorot and Bengio(2010)}]{glorot2010understanding}
\bibinfo{author}{Glorot, X.}, \bibinfo{author}{Bengio, Y.},
  \bibinfo{year}{2010}.
\newblock \bibinfo{title}{Understanding the difficulty of training deep
  feedforward neural networks}, in: \bibinfo{booktitle}{Proceedings of the
  thirteenth international conference on artificial intelligence and
  statistics}, pp. \bibinfo{pages}{249--256}.
\bibitem[{Gurcan et~al.(2009)Gurcan, Boucheron, Can, Madabhushi, Rajpoot and
  Yener}]{gurcan2009histopathological}
\bibinfo{author}{Gurcan, M.N.}, \bibinfo{author}{Boucheron, L.},
  \bibinfo{author}{Can, A.}, \bibinfo{author}{Madabhushi, A.},
  \bibinfo{author}{Rajpoot, N.}, \bibinfo{author}{Yener, B.},
  \bibinfo{year}{2009}.
\newblock \bibinfo{title}{Histopathological image analysis: A review}.
\newblock \bibinfo{journal}{IEEE reviews in biomedical engineering}
  \bibinfo{volume}{2}, \bibinfo{pages}{147}.
\bibitem[{Gutman et~al.(2013)Gutman, Cobb, Somanna, Park, Wang, Kurc, Saltz,
  Brat, Cooper and Kong}]{gutman2013cancer}
\bibinfo{author}{Gutman, D.A.}, \bibinfo{author}{Cobb, J.},
  \bibinfo{author}{Somanna, D.}, \bibinfo{author}{Park, Y.},
  \bibinfo{author}{Wang, F.}, \bibinfo{author}{Kurc, T.},
  \bibinfo{author}{Saltz, J.H.}, \bibinfo{author}{Brat, D.J.},
  \bibinfo{author}{Cooper, L.A.}, \bibinfo{author}{Kong, J.},
  \bibinfo{year}{2013}.
\newblock \bibinfo{title}{Cancer digital slide archive: an informatics resource
  to support integrated in silico analysis of tcga pathology data}.
\newblock \bibinfo{journal}{Journal of the American Medical Informatics
  Association} \bibinfo{volume}{20}, \bibinfo{pages}{1091--1098}.
\bibitem[{Huang et~al.(2017)Huang, Liu, Van Der~Maaten and
  Weinberger}]{huang2017densely}
\bibinfo{author}{Huang, G.}, \bibinfo{author}{Liu, Z.}, \bibinfo{author}{Van
  Der~Maaten, L.}, \bibinfo{author}{Weinberger, K.Q.}, \bibinfo{year}{2017}.
\newblock \bibinfo{title}{Densely connected convolutional networks}, in:
  \bibinfo{booktitle}{Proceedings of the IEEE conference on computer vision and
  pattern recognition}, pp. \bibinfo{pages}{4700--4708}.
\bibitem[{Janowczyk and Madabhushi(2016)}]{janowczyk2016deep}
\bibinfo{author}{Janowczyk, A.}, \bibinfo{author}{Madabhushi, A.},
  \bibinfo{year}{2016}.
\newblock \bibinfo{title}{Deep learning for digital pathology image analysis: A
  comprehensive tutorial with selected use cases}.
\newblock \bibinfo{journal}{Journal of pathology informatics}
  \bibinfo{volume}{7}.
\bibitem[{Jegou et~al.(2011)Jegou, Perronnin, Douze, S{\'a}nchez, Perez and
  Schmid}]{jegou2011aggregating}
\bibinfo{author}{Jegou, H.}, \bibinfo{author}{Perronnin, F.},
  \bibinfo{author}{Douze, M.}, \bibinfo{author}{S{\'a}nchez, J.},
  \bibinfo{author}{Perez, P.}, \bibinfo{author}{Schmid, C.},
  \bibinfo{year}{2011}.
\newblock \bibinfo{title}{Aggregating local image descriptors into compact
  codes}.
\newblock \bibinfo{journal}{IEEE transactions on pattern analysis and machine
  intelligence} \bibinfo{volume}{34}, \bibinfo{pages}{1704--1716}.
\bibitem[{Kalra et~al.(2019a)Kalra, Choi, Shah, Pantanowitz and
  Tizhoosh}]{kalra2019yottixel}
\bibinfo{author}{Kalra, S.}, \bibinfo{author}{Choi, C.}, \bibinfo{author}{Shah,
  S.}, \bibinfo{author}{Pantanowitz, L.}, \bibinfo{author}{Tizhoosh, H.},
  \bibinfo{year}{2019}a.
\newblock \bibinfo{title}{Yottixel--an image search engine for large archives
  of histopathology whole slide images}.
\newblock \bibinfo{journal}{arXiv preprint arXiv:1911.08748} .
\bibitem[{Kalra et~al.(2019b)Kalra, Tizhoosh, Shah, Choi, Damaskinos,
  Safarpoor, Shafiei, Babaie, Diamandis, Campbell et~al.}]{kalra2019pan}
\bibinfo{author}{Kalra, S.}, \bibinfo{author}{Tizhoosh, H.},
  \bibinfo{author}{Shah, S.}, \bibinfo{author}{Choi, C.},
  \bibinfo{author}{Damaskinos, S.}, \bibinfo{author}{Safarpoor, A.},
  \bibinfo{author}{Shafiei, S.}, \bibinfo{author}{Babaie, M.},
  \bibinfo{author}{Diamandis, P.}, \bibinfo{author}{Campbell, C.J.}, et~al.,
  \bibinfo{year}{2019}b.
\newblock \bibinfo{title}{Pan-cancer diagnostic consensus through searching
  archival histopathology images using artificial intelligence}.
\newblock \bibinfo{journal}{arXiv preprint arXiv:1911.08736} .
\bibitem[{Kather et~al.(2016)Kather, Weis, Bianconi, Melchers, Schad, Gaiser,
  Marx and Z{\"o}llner}]{kather2016multi}
\bibinfo{author}{Kather, J.N.}, \bibinfo{author}{Weis, C.A.},
  \bibinfo{author}{Bianconi, F.}, \bibinfo{author}{Melchers, S.M.},
  \bibinfo{author}{Schad, L.R.}, \bibinfo{author}{Gaiser, T.},
  \bibinfo{author}{Marx, A.}, \bibinfo{author}{Z{\"o}llner, F.G.},
  \bibinfo{year}{2016}.
\newblock \bibinfo{title}{Multi-class texture analysis in colorectal cancer
  histology}.
\newblock \bibinfo{journal}{Scientific reports} \bibinfo{volume}{6},
  \bibinfo{pages}{27988}.
\bibitem[{Kieffer et~al.(2017)Kieffer, Babaie, Kalra and
  Tizhoosh}]{kieffer2017convolutional}
\bibinfo{author}{Kieffer, B.}, \bibinfo{author}{Babaie, M.},
  \bibinfo{author}{Kalra, S.}, \bibinfo{author}{Tizhoosh, H.R.},
  \bibinfo{year}{2017}.
\newblock \bibinfo{title}{Convolutional neural networks for histopathology
  image classification: Training vs. using pre-trained networks}, in:
  \bibinfo{booktitle}{2017 Seventh International Conference on Image Processing
  Theory, Tools and Applications (IPTA)}, \bibinfo{organization}{IEEE}. pp.
  \bibinfo{pages}{1--6}.
\bibitem[{Kingma and Ba(2014)}]{kingma2014adam}
\bibinfo{author}{Kingma, D.P.}, \bibinfo{author}{Ba, J.}, \bibinfo{year}{2014}.
\newblock \bibinfo{title}{Adam: A method for stochastic optimization}.
\newblock \bibinfo{journal}{arXiv preprint arXiv:1412.6980} .
\bibitem[{Kraus et~al.(2017)Kraus, Grys, Ba, Chong, Frey, Boone and
  Andrews}]{kraus2017automated}
\bibinfo{author}{Kraus, O.Z.}, \bibinfo{author}{Grys, B.T.},
  \bibinfo{author}{Ba, J.}, \bibinfo{author}{Chong, Y.}, \bibinfo{author}{Frey,
  B.J.}, \bibinfo{author}{Boone, C.}, \bibinfo{author}{Andrews, B.J.},
  \bibinfo{year}{2017}.
\newblock \bibinfo{title}{Automated analysis of high-content microscopy data
  with deep learning}.
\newblock \bibinfo{journal}{Molecular systems biology} \bibinfo{volume}{13}.
\bibitem[{{Kumar} et~al.(2018){Kumar}, {Babaie} and {Tizhoosh}}]{8489574}
\bibinfo{author}{{Kumar}, M.D.}, \bibinfo{author}{{Babaie}, M.},
  \bibinfo{author}{{Tizhoosh}, H.R.}, \bibinfo{year}{2018}.
\newblock \bibinfo{title}{Deep barcodes for fast retrieval of histopathology
  scans}, in: \bibinfo{booktitle}{2018 International Joint Conference on Neural
  Networks (IJCNN)}, pp. \bibinfo{pages}{1--8}.
\newblock \DOIprefix\doi{10.1109/IJCNN.2018.8489574}.
\bibitem[{Kumar et~al.(2017)Kumar, Babaie, Zhu, Kalra and
  Tizhoosh}]{kumar2017comparative}
\bibinfo{author}{Kumar, M.D.}, \bibinfo{author}{Babaie, M.},
  \bibinfo{author}{Zhu, S.}, \bibinfo{author}{Kalra, S.},
  \bibinfo{author}{Tizhoosh, H.R.}, \bibinfo{year}{2017}.
\newblock \bibinfo{title}{A comparative study of cnn, bovw and lbp for
  classification of histopathological images}, in: \bibinfo{booktitle}{2017
  IEEE Symposium Series on Computational Intelligence (SSCI)},
  \bibinfo{organization}{IEEE}. pp. \bibinfo{pages}{1--7}.
\bibitem[{Lee et~al.(2017)Lee, Jung, Agarwal and Kim}]{lee2017can}
\bibinfo{author}{Lee, H.S.}, \bibinfo{author}{Jung, H.},
  \bibinfo{author}{Agarwal, A.A.}, \bibinfo{author}{Kim, J.},
  \bibinfo{year}{2017}.
\newblock \bibinfo{title}{Can deep neural networks match the related objects?:
  A survey on imagenet-trained classification models}.
\newblock \bibinfo{journal}{arXiv preprint arXiv:1709.03806} .
\bibitem[{Liu et~al.(2018)Liu, Ouyang, Wang, Fieguth, Chen, Liu and
  Pietik{\"a}inen}]{liu2018deep}
\bibinfo{author}{Liu, L.}, \bibinfo{author}{Ouyang, W.}, \bibinfo{author}{Wang,
  X.}, \bibinfo{author}{Fieguth, P.}, \bibinfo{author}{Chen, J.},
  \bibinfo{author}{Liu, X.}, \bibinfo{author}{Pietik{\"a}inen, M.},
  \bibinfo{year}{2018}.
\newblock \bibinfo{title}{Deep learning for generic object detection: A
  survey}.
\newblock \bibinfo{journal}{arXiv preprint arXiv:1809.02165} .
\bibitem[{Liu et~al.(2017)Liu, Gadepalli, Norouzi, Dahl, Kohlberger, Boyko,
  Venugopalan, Timofeev, Nelson, Corrado et~al.}]{liu2017detecting}
\bibinfo{author}{Liu, Y.}, \bibinfo{author}{Gadepalli, K.},
  \bibinfo{author}{Norouzi, M.}, \bibinfo{author}{Dahl, G.E.},
  \bibinfo{author}{Kohlberger, T.}, \bibinfo{author}{Boyko, A.},
  \bibinfo{author}{Venugopalan, S.}, \bibinfo{author}{Timofeev, A.},
  \bibinfo{author}{Nelson, P.Q.}, \bibinfo{author}{Corrado, G.S.}, et~al.,
  \bibinfo{year}{2017}.
\newblock \bibinfo{title}{Detecting cancer metastases on gigapixel pathology
  images}.
\newblock \bibinfo{journal}{arXiv preprint arXiv:1703.02442} .
\bibitem[{Madabhushi(2009)}]{madabhushi2009digital}
\bibinfo{author}{Madabhushi, A.}, \bibinfo{year}{2009}.
\newblock \bibinfo{title}{Digital pathology image analysis: opportunities and
  challenges}.
\newblock \bibinfo{journal}{Imaging in medicine} \bibinfo{volume}{1},
  \bibinfo{pages}{7}.
\bibitem[{Mormont et~al.(2018)Mormont, Geurts and
  Mar{\'e}e}]{mormont2018comparison}
\bibinfo{author}{Mormont, R.}, \bibinfo{author}{Geurts, P.},
  \bibinfo{author}{Mar{\'e}e, R.}, \bibinfo{year}{2018}.
\newblock \bibinfo{title}{Comparison of deep transfer learning strategies for
  digital pathology}, in: \bibinfo{booktitle}{Proceedings of the IEEE
  Conference on Computer Vision and Pattern Recognition Workshops}, pp.
  \bibinfo{pages}{2262--2271}.
\bibitem[{Najafabadi et~al.(2015)Najafabadi, Villanustre, Khoshgoftaar, Seliya,
  Wald and Muharemagic}]{najafabadi2015deep}
\bibinfo{author}{Najafabadi, M.M.}, \bibinfo{author}{Villanustre, F.},
  \bibinfo{author}{Khoshgoftaar, T.M.}, \bibinfo{author}{Seliya, N.},
  \bibinfo{author}{Wald, R.}, \bibinfo{author}{Muharemagic, E.},
  \bibinfo{year}{2015}.
\newblock \bibinfo{title}{Deep learning applications and challenges in big data
  analytics}.
\newblock \bibinfo{journal}{Journal of Big Data} \bibinfo{volume}{2},
  \bibinfo{pages}{1}.
\bibitem[{Nanni et~al.(2018)Nanni, Ghidoni and Brahnam}]{nanni2018ensemble}
\bibinfo{author}{Nanni, L.}, \bibinfo{author}{Ghidoni, S.},
  \bibinfo{author}{Brahnam, S.}, \bibinfo{year}{2018}.
\newblock \bibinfo{title}{Ensemble of convolutional neural networks for
  bioimage classification}.
\newblock \bibinfo{journal}{Applied Computing and Informatics} .
\bibitem[{Niazi et~al.(2019)Niazi, Parwani and Gurcan}]{niazi2019digital}
\bibinfo{author}{Niazi, M.K.K.}, \bibinfo{author}{Parwani, A.V.},
  \bibinfo{author}{Gurcan, M.N.}, \bibinfo{year}{2019}.
\newblock \bibinfo{title}{Digital pathology and artificial intelligence}.
\newblock \bibinfo{journal}{The Lancet Oncology} \bibinfo{volume}{20},
  \bibinfo{pages}{e253--e261}.
\bibitem[{Onder et~al.(2014)Onder, Zengin and Sarioglu}]{onder2014review}
\bibinfo{author}{Onder, D.}, \bibinfo{author}{Zengin, S.},
  \bibinfo{author}{Sarioglu, S.}, \bibinfo{year}{2014}.
\newblock \bibinfo{title}{A review on color normalization and color
  deconvolution methods in histopathology}.
\newblock \bibinfo{journal}{Applied Immunohistochemistry \& Molecular
  Morphology} \bibinfo{volume}{22}, \bibinfo{pages}{713--719}.
\bibitem[{Phan et~al.(2016)Phan, Kumar, Kim and Feng}]{phan2016transfer}
\bibinfo{author}{Phan, H.T.H.}, \bibinfo{author}{Kumar, A.},
  \bibinfo{author}{Kim, J.}, \bibinfo{author}{Feng, D.}, \bibinfo{year}{2016}.
\newblock \bibinfo{title}{Transfer learning of a convolutional neural network
  for hep-2 cell image classification}, in: \bibinfo{booktitle}{2016 IEEE 13th
  International Symposium on Biomedical Imaging (ISBI)},
  \bibinfo{organization}{IEEE}. pp. \bibinfo{pages}{1208--1211}.
\bibitem[{Raghu et~al.(2019)Raghu, Zhang, Kleinberg and
  Bengio}]{raghu2019transfusion}
\bibinfo{author}{Raghu, M.}, \bibinfo{author}{Zhang, C.},
  \bibinfo{author}{Kleinberg, J.}, \bibinfo{author}{Bengio, S.},
  \bibinfo{year}{2019}.
\newblock \bibinfo{title}{Transfusion: Understanding transfer learning for
  medical imaging}, in: \bibinfo{booktitle}{Advances in Neural Information
  Processing Systems}, pp. \bibinfo{pages}{3342--3352}.
\bibitem[{Rajpurkar et~al.(2017)Rajpurkar, Irvin, Zhu, Yang, Mehta, Duan, Ding,
  Bagul, Langlotz, Shpanskaya et~al.}]{rajpurkar2017chexnet}
\bibinfo{author}{Rajpurkar, P.}, \bibinfo{author}{Irvin, J.},
  \bibinfo{author}{Zhu, K.}, \bibinfo{author}{Yang, B.},
  \bibinfo{author}{Mehta, H.}, \bibinfo{author}{Duan, T.},
  \bibinfo{author}{Ding, D.}, \bibinfo{author}{Bagul, A.},
  \bibinfo{author}{Langlotz, C.}, \bibinfo{author}{Shpanskaya, K.}, et~al.,
  \bibinfo{year}{2017}.
\newblock \bibinfo{title}{Chexnet: Radiologist-level pneumonia detection on
  chest x-rays with deep learning}.
\newblock \bibinfo{journal}{arXiv preprint arXiv:1711.05225} .
\bibitem[{Sharif~Razavian et~al.(2014)Sharif~Razavian, Azizpour, Sullivan and
  Carlsson}]{sharif2014cnn}
\bibinfo{author}{Sharif~Razavian, A.}, \bibinfo{author}{Azizpour, H.},
  \bibinfo{author}{Sullivan, J.}, \bibinfo{author}{Carlsson, S.},
  \bibinfo{year}{2014}.
\newblock \bibinfo{title}{Cnn features off-the-shelf: an astounding baseline
  for recognition}, in: \bibinfo{booktitle}{Proceedings of the IEEE conference
  on computer vision and pattern recognition workshops}, pp.
  \bibinfo{pages}{806--813}.
\bibitem[{Shin et~al.(2016)Shin, Roth, Gao, Lu, Xu, Nogues, Yao, Mollura and
  Summers}]{shin2016deep}
\bibinfo{author}{Shin, H.C.}, \bibinfo{author}{Roth, H.R.},
  \bibinfo{author}{Gao, M.}, \bibinfo{author}{Lu, L.}, \bibinfo{author}{Xu,
  Z.}, \bibinfo{author}{Nogues, I.}, \bibinfo{author}{Yao, J.},
  \bibinfo{author}{Mollura, D.}, \bibinfo{author}{Summers, R.M.},
  \bibinfo{year}{2016}.
\newblock \bibinfo{title}{Deep convolutional neural networks for computer-aided
  detection: Cnn architectures, dataset characteristics and transfer learning}.
\newblock \bibinfo{journal}{IEEE transactions on medical imaging}
  \bibinfo{volume}{35}, \bibinfo{pages}{1285--1298}.
\bibitem[{{Spanhol} et~al.(2017){Spanhol}, {Oliveira}, {Cavalin}, {Petitjean}
  and {Heutte}}]{8122889}
\bibinfo{author}{{Spanhol}, F.A.}, \bibinfo{author}{{Oliveira}, L.S.},
  \bibinfo{author}{{Cavalin}, P.R.}, \bibinfo{author}{{Petitjean}, C.},
  \bibinfo{author}{{Heutte}, L.}, \bibinfo{year}{2017}.
\newblock \bibinfo{title}{Deep features for breast cancer histopathological
  image classification}, in: \bibinfo{booktitle}{2017 IEEE International
  Conference on Systems, Man, and Cybernetics (SMC)}, pp.
  \bibinfo{pages}{1868--1873}.
\newblock \DOIprefix\doi{10.1109/SMC.2017.8122889}.
\bibitem[{Srivastava et~al.(2015)Srivastava, Greff and
  Schmidhuber}]{srivastava2015training}
\bibinfo{author}{Srivastava, R.K.}, \bibinfo{author}{Greff, K.},
  \bibinfo{author}{Schmidhuber, J.}, \bibinfo{year}{2015}.
\newblock \bibinfo{title}{Training very deep networks}, in:
  \bibinfo{booktitle}{Advances in neural information processing systems}, pp.
  \bibinfo{pages}{2377--2385}.
\bibitem[{{Sun} et~al.(2019){Sun}, {Zeng}, {Xu}, {Peng} and {Ma}}]{8854180}
\bibinfo{author}{{Sun}, H.}, \bibinfo{author}{{Zeng}, X.},
  \bibinfo{author}{{Xu}, T.}, \bibinfo{author}{{Peng}, G.},
  \bibinfo{author}{{Ma}, Y.}, \bibinfo{year}{2019}.
\newblock \bibinfo{title}{Computer-aided diagnosis in histopathological images
  of the endometrium using a convolutional neural network and attention
  mechanisms}.
\newblock \bibinfo{journal}{IEEE Journal of Biomedical and Health Informatics}
  , \bibinfo{pages}{1--1}\DOIprefix\doi{10.1109/JBHI.2019.2944977}.
\bibitem[{Tizhoosh and Pantanowitz(2018)}]{tizhoosh2018artificial}
\bibinfo{author}{Tizhoosh, H.R.}, \bibinfo{author}{Pantanowitz, L.},
  \bibinfo{year}{2018}.
\newblock \bibinfo{title}{Artificial intelligence and digital pathology:
  Challenges and opportunities}.
\newblock \bibinfo{journal}{Journal of pathology informatics}
  \bibinfo{volume}{9}.
\bibitem[{Tizhoosh et~al.(2016)Tizhoosh, Zhu, Lo, Chaudhari and
  Mehdi}]{tizhoosh2016minmax}
\bibinfo{author}{Tizhoosh, H.R.}, \bibinfo{author}{Zhu, S.},
  \bibinfo{author}{Lo, H.}, \bibinfo{author}{Chaudhari, V.},
  \bibinfo{author}{Mehdi, T.}, \bibinfo{year}{2016}.
\newblock \bibinfo{title}{Minmax radon barcodes for medical image retrieval},
  in: \bibinfo{booktitle}{International Symposium on Visual Computing},
  \bibinfo{organization}{Springer}. pp. \bibinfo{pages}{617--627}.
\bibitem[{Tomczak et~al.(2015)Tomczak, Czerwi{\'n}ska and
  Wiznerowicz}]{tomczak2015cancer}
\bibinfo{author}{Tomczak, K.}, \bibinfo{author}{Czerwi{\'n}ska, P.},
  \bibinfo{author}{Wiznerowicz, M.}, \bibinfo{year}{2015}.
\newblock \bibinfo{title}{The cancer genome atlas (tcga): an immeasurable
  source of knowledge}.
\newblock \bibinfo{journal}{Contemporary oncology} \bibinfo{volume}{19},
  \bibinfo{pages}{A68}.
\bibitem[{Travis(2014)}]{travis2014pathology}
\bibinfo{author}{Travis, W.D.}, \bibinfo{year}{2014}.
\newblock \bibinfo{title}{Pathology and diagnosis of neuroendocrine tumors:
  lung neuroendocrine}.
\newblock \bibinfo{journal}{Thoracic surgery clinics} \bibinfo{volume}{24},
  \bibinfo{pages}{257--266}.
\bibitem[{Watanabe et~al.(2017)Watanabe, Hori, Le~Roux and
  Hershey}]{watanabe2017student}
\bibinfo{author}{Watanabe, S.}, \bibinfo{author}{Hori, T.},
  \bibinfo{author}{Le~Roux, J.}, \bibinfo{author}{Hershey, J.R.},
  \bibinfo{year}{2017}.
\newblock \bibinfo{title}{Student-teacher network learning with enhanced
  features}, in: \bibinfo{booktitle}{2017 IEEE International Conference on
  Acoustics, Speech and Signal Processing (ICASSP)},
  \bibinfo{organization}{IEEE}. pp. \bibinfo{pages}{5275--5279}.
\bibitem[{Wei et~al.(2019)Wei, Tafe, Linnik, Vaickus, Tomita and
  Hassanpour}]{wei2019pathologist}
\bibinfo{author}{Wei, J.W.}, \bibinfo{author}{Tafe, L.J.},
  \bibinfo{author}{Linnik, Y.A.}, \bibinfo{author}{Vaickus, L.J.},
  \bibinfo{author}{Tomita, N.}, \bibinfo{author}{Hassanpour, S.},
  \bibinfo{year}{2019}.
\newblock \bibinfo{title}{Pathologist-level classification of histologic
  patterns on resected lung adenocarcinoma slides with deep neural networks}.
\newblock \bibinfo{journal}{Scientific reports} \bibinfo{volume}{9},
  \bibinfo{pages}{3358}.
\bibitem[{Yu et~al.(2016)Yu, Deng, Seide and Li}]{yu2016discriminative}
\bibinfo{author}{Yu, D.}, \bibinfo{author}{Deng, L.}, \bibinfo{author}{Seide,
  F.T.B.}, \bibinfo{author}{Li, G.}, \bibinfo{year}{2016}.
\newblock \bibinfo{title}{Discriminative pretraining of deep neural networks}.
\newblock \bibinfo{note}{US Patent 9,235,799}.
\bibitem[{Zhang et~al.(2019a)Zhang, Lu, Li, Kim and Wang}]{zhang2019full}
\bibinfo{author}{Zhang, J.}, \bibinfo{author}{Lu, C.}, \bibinfo{author}{Li,
  X.}, \bibinfo{author}{Kim, H.J.}, \bibinfo{author}{Wang, J.},
  \bibinfo{year}{2019}a.
\newblock \bibinfo{title}{A full convolutional network based on densenet for
  remote sensing scene classification}.
\newblock \bibinfo{journal}{Math. Biosci. Eng} \bibinfo{volume}{16},
  \bibinfo{pages}{3345--3367}.
\bibitem[{Zhang et~al.(2019b)Zhang, Guo, Wang, Yuan and
  Ding}]{zhang2019multiple}
\bibinfo{author}{Zhang, K.}, \bibinfo{author}{Guo, Y.}, \bibinfo{author}{Wang,
  X.}, \bibinfo{author}{Yuan, J.}, \bibinfo{author}{Ding, Q.},
  \bibinfo{year}{2019}b.
\newblock \bibinfo{title}{Multiple feature reweight densenet for image
  classification}.
\newblock \bibinfo{journal}{IEEE Access} \bibinfo{volume}{7},
  \bibinfo{pages}{9872--9880}.
\bibitem[{Zhu et~al.(2018)Zhu, Li, Kalra and Tizhoosh}]{zhu2018multiple}
\bibinfo{author}{Zhu, S.}, \bibinfo{author}{Li, Y.}, \bibinfo{author}{Kalra,
  S.}, \bibinfo{author}{Tizhoosh, H.R.}, \bibinfo{year}{2018}.
\newblock \bibinfo{title}{Multiple disjoint dictionaries for representation of
  histopathology images}.
\newblock \bibinfo{journal}{Journal of Visual Communication and Image
  Representation} \bibinfo{volume}{55}, \bibinfo{pages}{243--252}.

\end{thebibliography}
